\documentclass[aps,floatfix,twocolumn,preprintnumbers,sort&compress,square,numbers,nofootinbib,pra]{revtex4-2}
\usepackage{graphicx}  
\usepackage{multirow}

\usepackage{fancyhdr}
\usepackage{longtable}
\usepackage{parskip}
\usepackage{pgfplots}
\usepackage[T1]{fontenc}
\usepackage{dcolumn}   

\usepackage{bm}        
\usepackage{amsfonts}  
\usepackage{amsmath}   
\usepackage{amssymb}   
\usepackage{mathtools}
\usepackage{array}
\usepackage{braket}
\usepackage{breqn}
\usepackage{bbold}
\usepackage{booktabs}

\usepackage{hyperref}
\usepackage[capitalise]{cleveref}
\usepackage{float}
\usepackage{standalone}
\usepackage[caption=false]{subfig}
\usepackage{tikz}
\usetikzlibrary{decorations.pathreplacing}
\usetikzlibrary{calc}

\pgfplotsset{compat=1.17}

\newcommand{\I}{\imath}

\newcommand{\pwisein}{\left\{ \begin{array}{ll}}
\newcommand{\pwiseout}{\end{array}\right.}
\newcommand{\bracket}[2]{\left\langle #1 | #2 \right\rangle}

\newcommand\Tstrut{\rule{0pt}{2.8ex}}         
\newcommand\Bstrut{\rule[-1.7ex]{0pt}{0pt}}   
\newcommand\TBstrut{\Tstrut\Bstrut}           

\newcommand{\foo}[1]{%
\begin{tikzpicture}[#1][line cap = round, line join = round, >=triangle 45]
		\tikzstyle{point1}=[circle, draw=black, inner sep=0.03cm]
		\tikzstyle{point2}=[circle, draw=black, inner sep=0.03cm]
		\tikzstyle{point3}=[circle, draw=black, inner sep=0.03cm]
		\node[point1] (origin) at (0,0) {$\small 1$};
		\node[point2] (A) at (1,0.1)   {$\small 2$};
		\node[point3] (B) at (0.3, 0.7) {$\small 3$};
		\coordinate (M) at ($ (origin) !.5! (A) $);
		
		\draw[-stealth,color=black] (origin) -- (A) node[pos=0.5, above]{};
		\draw[-stealth,color=black] (A) -- (B) node[pos=0.5, left]{};
		\draw[-stealth,color=black] (B) -- (origin) node[pos=0.5, right]{};
\end{tikzpicture}%
}

\newcommand{\fooo}[1]{%
\begin{tikzpicture}[#1][line cap = round, line join = round, >=triangle 45]
		\tikzstyle{point1}=[circle, draw=black, inner sep=0.03cm]
		\tikzstyle{point2}=[circle, draw=black, inner sep=0.03cm]
		\tikzstyle{point3}=[circle, draw=black, inner sep=0.03cm]
		\node[point1] (origin) at (0,0) {$\small 2$};
		\node[point2] (A) at (1,0.1)   {$\small 1$};
		\node[point3] (B) at (0.3, 0.7) {$\small 3$};
		\coordinate (M) at ($ (origin) !.5! (A) $);
		
		\draw[-stealth,color=black] (origin) -- (A) node[pos=0.5, above]{};
		\draw[-stealth,color=black] (A) -- (B) node[pos=0.5, left]{};
		\draw[-stealth,color=black] (B) -- (origin) node[pos=0.5, right]{};
\end{tikzpicture}%
}

\newcommand{\foooo}[1]{%
\begin{tikzpicture}[#1][line cap = round, line join = round, >=triangle 45]
		\tikzstyle{point1}=[circle, draw=black, inner sep=0.03cm]
		\tikzstyle{point2}=[circle, draw=black, inner sep=0.03cm]
		\tikzstyle{point3}=[circle, draw=black, inner sep=0.03cm]
		\node[point1] (origin) at (0,0) {$\small 2$};
		\node[point2] (A) at (1,0.1)   {$\small 3$};
		\node[point3] (B) at (0.3, 0.7) {$\small 1$};
		\coordinate (M) at ($ (origin) !.5! (A) $);
		
		\draw[-stealth,color=black] (origin) -- (A) node[pos=0.5, above]{};
		\draw[-stealth,color=black] (A) -- (B) node[pos=0.5, left]{};
		\draw[-stealth,color=black] (B) -- (origin) node[pos=0.5, right]{};
\end{tikzpicture}%
}

\makeatletter
\newsavebox{\@brx}
\newcommand{\llangle}[1][]{\savebox{\@brx}{\(\m@th{#1\langle}\)}%
	\mathopen{\copy\@brx\mkern2mu\kern-0.9\wd\@brx\usebox{\@brx}}}
\newcommand{\rrangle}[1][]{\savebox{\@brx}{\(\m@th{#1\rangle}\)}%
	\mathclose{\copy\@brx\mkern2mu\kern-0.9\wd\@brx\usebox{\@brx}}}
\makeatother

\begin{document}
\title{A multichannel hyperspherical model for Efimov physics with van der Waals interactions controlled by a Feshbach resonance}

\author{Kajsa--My Tempest}
\author{Svante Jonsell}
\affiliation{Department of Physics Stockholm University\char`,{} SE-10691 Stockholm\char`,{} Sweden}

\date{\today}

\begin{abstract}  
Here we present a four-channel model that incorporates a magnetically tunable Feshbach resonance in a system of three atoms that interact via pairwise van der Waals interactions. Our method is designed to model recent experiments where the tunability of the scattering length has been used to study three-body Efimov states, which appear in the limit of a diverging two-body scattering length. Using this model, we calculate three-body adiabatic and effective potential curves and study how the strength (or width) of the Feshbach resonance affects the three-body potential that is connected to the Efimov effect. We find that the position of the repulsive barrier, which has been used to explain the so-called van der Waals universality in broad resonances, is slightly shifted as the narrow resonance limit is approached and that this shift is correlated to the appearance of two avoided crossings in the adiabatic energy landscape. More importantly, the attractive well is markedly shifted upward in energy and is extremely shallow for the narrowest resonance. We argue that this behavior is connected to the breakdown of van der Waals universality for weak (narrow) resonances.
\end{abstract}

\maketitle 

\section{Introduction}

Near resonant two-body forces are famously known to give rise to a series of energy levels in the quantum three-body spectrum. Exactly on resonance, an infinite number of these levels appear obeying a discrete geometric scaling law. These energy levels are called Efimov states \cite{Efimov:1970zz} and they have a number of interesting universal properties. The term Efimov physics \cite{Naidon_2017} refers to a range of universal phenomena that arise in few-body systems displaying this effect.

Despite many efforts, the experimental confirmation of Efimov's result proved challenging. The reason is that the ideal Efimov scenario requires that the parameter which characterizes the resonant two-body interaction, the  $s$-wave scattering length $a$,  is much larger in magnitude than the pairwise interaction range. 

In experimental settings, this condition is rarely perfectly fulfilled. An exception is  ultracold atomic clouds where the scattering length can be controlled by means of a magnetically tunable Feshbach resonance \cite{Inouye98}. A Feshbach resonance occurs when two colliding atoms weakly couple to a quasibound molecular state (which in the absence of this coupling would be strictly bound). 
For example, a pair of atoms with two different spin configurations may couple through hyperfine interactions. If the two molecular states have different magnetic moments, an applied magnetic field will shift their relative energies through the Zeeman effect. The position of the closed channel bound state can thus be controlled by tuning the strength of the magnetic field. 

In the ultracold regime, collisions occur in the zero energy limit. At some magnetic field value $B_0$ the quasibound energy level of the closed-channel molecule will coincide with the threshold energy of the open channel. This results in resonant coupling in the limit of zero-energy scattering, and hence the $s$-wave scattering length $a(B)$ diverges. In the vicinity of this magnetic field, the magnetic-field dependence of the scattering length is given by \cite{PhysRevA.51.4852}
\begin{equation}\label{eq:Bdep_a}
    a(B)=a_{\mathrm{bg}}\left(1-\frac{\Delta B}{B-B_0}\right),
\end{equation}
where $a_{\mathrm{bg}}$ is the off-resonant background scattering length and $\Delta B$ is the width of the resonance. The theory behind Feshbach resonances and their classification will be discussed further in \cref{sec:Feshbach_resonances}. For the moment, suffice it to say that they are classified as either strong (entrance-channel dominated) or weak (closed-channel dominated), depending on their width and the parameters of the two-body interaction potential.

Using this technique, the appearance of near-threshold Efimov states in a cloud of ultracold caesium atoms was detected through an enhanced rate of three-body recombination at a certain strength of the magnetic field \cite{Kraemer2006}. Since then Feshbach tuning of the scattering length has been used to search for Efimov states in a number of alkali systems, and such states have been reported both in systems of three identical bosons: $^7$Li \cite{Gross2009,Pollack2009}, $^{39}$K \cite{Zaccanti2009,Chapurin2019,Xie2020}, $^{85}$Rb \cite{Wild2012}, as well as in mixed boson or boson-fermion systems: $^{41}$K--$^{87}$Rb \cite{Barontini2009,Wacker2016},$^{40}$K--$^{87}$Rb \cite{Bloom2013} $^7$Li--$^{87}$Rb \cite{Maier2015}, $^6$Li--$^{133}$Cs \cite{Tung2014,Pires2014}, and even in the purely fermionic system:  $^6$Li \cite{Williams2009,Wenz2009,Lompe2010,Huang2014b} (using a mixture of different hyperfine states).

Notably, the Efimov effect has also been found in the Helium trimer, a system without Feshbach resonances of the type discussed above \cite{Kunitski2015}.

While Efimov's theory predicts a universal scaling between the parameters of successive Efimov states, the absolute scales of these remain nonuniversal, i.e., set by the details of the atom-atom interaction and generally different for different atoms. However, as experimental observations of Efimov states accumulated, an unexpected additional universality was revealed. For strong resonances, the Efimov states appeared at magnetic fields where $a\approx-9 r_{\mathrm{vdW}}$, with $r_{\mathrm{vdW}}$ being the van der Waals length \cite{Berninger2011}. This phenomenon is called van der Waals universality and its origin has been explained by the suppression of the pair correlations at short distances, which manifests itself in the form of a repulsive barrier at a specific position in three-body effective potentials \cite{Wang_2012_Origin}.

The theoretical analysis in \cite{Wang_2012_Origin}, like many theoretical works on Efimov physics, uses a single-channel model based on a model potential. Here a van der Waals potential tuned so that $a\rightarrow\pm \infty$ was used. Thus the physics of the Feshbach resonance is only included through the single parameter $a(B)$ taken at $B=B_0$. Therefore the model per definition cannot investigate any effects depending on the strength or width $\Delta B$ of the resonance, which depends on the size of the coupling between the initial open channel and the molecular closed channel, or any other effects related to the multi-channel nature of the resonant collision physics.

Only a small number of theoretical works explicitly allowing for the multi-channel nature of Feshbach resonances have been published:
Some qualitative aspects have been investigated using methods based on the coordinate-space Faddeev equations with zero-range interactions \cite{Jonsell2004,Metha,Sorensen2013,Sorensen2012}, or using methods based on creation and annihilation operators \cite{Gogolin,Jona,Yudkin}. In contrast, quantitative agreement with experiments using $^{133}$Cs \cite{YWang2014} and $^{39}$K \cite{Xie2020} have been obtained by solving the Schr\"odinger equation in the adiabatic hyperspherical approximation using a Lennard--Jones potential with experimentally determined parameters. 
Recent work has addressed  physics similar to the present paper, but using a rather different approach based on the momentum-space Faddeev equations and using a separable interaction \cite{Secker2021,Kraats_Kookelmans}.

With a model that can describe closed-channel dominated resonances, it is possible to examine how the width of the resonance affects the three-body spectrum and its universal properties. More specifically, it makes it possible to numerically investigate the limit of the van der Waals universality \cite{Wang_2012_Origin,Mestrom_2017}. In experiments, the van der Waals universality appears in the broad and intermediate resonance regime \cite{Roy2013,Johansen2017}. However, for narrow resonances there are indications that this universality breaks down.

When generalized to three atoms, a two-channel model for binary interactions requires four channels, since there are three different ways to pair the atoms in the closed channel. In this paper, we present a method for solving the problem of three identical bosons interacting via pairwise van der Waals interactions using a four-channel model that incorporates a magnetically tunable Feshbach resonance. An advantage of this model is that it makes it possible to study various aspects of Efimov physics beyond the single-channel approximation and probe the closed-channel dominated narrow resonance regime. 

Most experimentally studied Efimov states have utilized Feshbach resonances with intermediate to strong coupling, undoubtedly because weakly coupled resonances usually require extremely precise magnetic-field control, which presents experimental challenges especially if $B_0$ is large. For our numerical results, we have used parameters from narrow resonances in $^{23} \rm Na$. To our knowledge, no attempt has been made to search for Efimov states at these resonances. The choice was instead motivated by the fact that these resonances represent clean examples of weak to extremely weak resonances, and are very well understood from theory \cite{Mies}. We expect that the qualitative features found here will carry over to similar resonances in other systems.

In this work, we examine the numerical three-body hyperradial potential curves obtained using our four-channel model, and compare them to single-channel calculations. We find several features depending on the resonance strength. Our prime finding is the appearance of additional avoided level crossings in the adiabatic potential energy landscape as the width of the resonance is decreased. We discuss how these avoided crossings affect the diabatic Efimov potential and analyze the subsequent effects on features related to van der Waals universality. 

\section{Two-Body Scattering with a Feshbach Resonance}\label{sec:Feshbach_resonances}

The description of a three-body scattering process in the presence of a Feshbach resonance requires a small set of parameters obtained from the equivalent two-body scattering calculations. For our three-body model we have considered the scattering of three bosonic sodium atoms and used the two-body model described in \cite{Mies} to extract the necessary parameters. Here we briefly describe the two-body scattering picture by closely following the theory presented in \cite{Mies,Nygaard,Kohler,Chin_2010}.

\subsection{Coupled channel scattering}

Magnetically tunable resonances may appear in atoms where a non-zero nuclear spin $\vec{i}$ couples to a non-zero electronic angular momentum $\vec{j}$ to form a total (atomic) angular momentum $\vec{f}=\vec{i}+\vec{j}$. Examples are given by the elements in Group I of the periodic table, with ground-state terms  $\prescript{2}{}{\mathrm{S}}^{}_{1/2}$. These atoms all have vanishing orbital angular momentum, giving $\vec{j}=\vec{s}$. The energy of the $\prescript{2}{}{\mathrm{S}}^{}_{1/2}$ level is thus split into two hyperfine levels with the corresponding quantum numbers $f=i-1/2$ and $f=i+1/2$. The difference in energy of these two levels is the hyperfine splitting $E_{\mathrm{hf}}$. In the presence of an external magnetic field, the Zeeman interaction couples the spin of the electron and the nucleus to the magnetic field, which results in a splitting of the hyperfine levels into its Zeeman levels, according to their projections $m_f\hbar$ of $\vec{f}$ along the magnetic field axis. 

We consider the scattering of two $\prescript{2}{}{\mathrm{S}}^{}_{1/2}$ $\mathrm{Na}$ atoms with nuclear spin $i=3/2$, resulting in the possible hyperfine levels $f=1$ and $f=2$, with Zeeman levels (in ascending energy order) $m_{f=1}=1,0,-1$ and $m_{f=2}=-2,-1,0,1,2$. These states are labeled $\ket{a}, \ket{b}, \ldots, \ket{h}$. The spin-projection $m_s$ onto the magnetic field  equals $-1/2$ for the four lowest Zeeman levels and $1/2$ for the four highest levels (see Figure 1 in \cite{Mies}).

In an ultracold collision, the hyperfine and Zeeman energies are large compared to the kinetic energy of the atoms, and in a sufficiently strong magnetic field, the Zeeman interaction dominates. The scattering channels are then classified according to the asymptotic Zeeman levels of each atom, that is $\alpha\beta=\{ f_1m_{f1} f_2m_{f2}\}$, where the curly brackets indicate bosonic symmetry. When all channels are closed, the total energy of the system $E$ will be associated with a strictly bound molecule. However, if the state couples to at least one open channel, $E$ will be associated with a discrete state in a closed channel embedded in the scattering continuum, or in other words a resonance.

Furthermore, the restriction to ultracold collisions means that we need only consider $s$-wave scattering, i.e., to a good approximation the relative angular momentum between two colliding atoms can be ignored. Thus the 
magnetic quantum number is the sum of the contributions from the two atoms, $M_F=m_{f1}+ m_{f2}$. Owing to the invariance under rotations around the magnetic-field axis (taken to be along the $z$-axis), $M_F$ is conserved during a collision. As the atoms approach each other they will reach a point where the electrons uncouple from the nuclei and recouple to each other, forming a molecular electronic spin $\vec{S}=\vec{s}_1+\vec{s}_2$, resulting in either a molecular singlet ($S=M_S=0$) or a triplet ($S=1, M_S=0,\pm1$). The hyperfine interaction weakly couples states with different electronic spins, but the same total spin $M_F$.

While $M_S$ is not strictly conserved, the magnetic moment of a molecular channel is approximately given by $\mu\simeq \mu_B M_S$. This makes it possible to use $B$ as a knob to control the energy separation between the asymptotic initial state and a closed channel state through their difference $\delta\mu$ in magnetic moments. At the point of recoupling the intrinsic couplings mix the channels, making it possible for the atom pair to transition from the initial channel to a closed channel, where the atoms become temporarily bound.

\begin{table}[t]
\centering
\begin{tabular*}{1.0\linewidth}{@{\extracolsep{\fill}}cccccc}
\hline
\hline
$f_1 m_{f_1}$ & $m_{s_1}$ & $f_2 m_{f_2}$ & $m_{s_2}$ & $\alpha\beta$ & $M_S$\TBstrut\\
\hline
$1\,1$ & $-\frac{1}{2}$ & $1\,1$ & $-\frac{1}{2}$ & $aa$ & $-1$\Tstrut \\
$1\,1$ & $-\frac{1}{2}$ & $2\,1$ & $+\frac{1}{2}$ & $ag$ & $0$ \\
$1\,0$ & $-\frac{1}{2}$ & $2\,2$ & $+\frac{1}{2}$ & $bh$ & $0$ \\
$2\,0$ & $+\frac{1}{2}$ & $2\,2$ & $+\frac{1}{2}$ & $fh$ & $1$ \\
$2\,1$ & $+\frac{1}{2}$ & $2\,1$ & $+\frac{1}{2}$ & $gg$ & $1$\Bstrut \\
\hline
\hline
\end{tabular*}
\caption{The table shows the quantum numbers for the different $s$-wave scattering channels that have $M_F=2$, along with a labeling convention. The channels are listed in order of increasing internal energy $E_{\alpha\beta}$.}
\label{table:Mchannels}
\end{table}
We assume that atoms are trapped in their lowest Zeeman state, which means $aa$ is the entrance channel. This state has $M_F=2$, and can couple to four additional combinations of atomic states that result in  molecular states ($\ket{\{\alpha\beta\}}$) with the same value of $M_F$, see \cref{table:Mchannels}. As long as the kinetic energy is less than the Zeeman splitting, only the $aa$ channel is open asymptotically and it couples to the four asymptotically closed channels at shorter $r$. The collision problem can then be described by the coupled-channel
method \cite{Ahn,Stoof,Moerdijk}, in which the total scattering state is expanded into a sum of individual channel states  
\begin{equation}
    \Psi=\sum_{\alpha\beta}\psi_{\alpha\beta}(\vec{r})\ket{\{\alpha\beta\}},
    \label{eq:channelexpansion}
\end{equation}
where $\ket{\{\alpha\beta\}}=\ket{\alpha}\otimes\ket{\beta}$ asymptotically (i.e., as $r\rightarrow\infty$). The expansion \cref{eq:channelexpansion} is substituted into the Schr{\"o}dinger equation, where the dynamics of the system is described by an effective Hamiltonian consisting of $H_0$ with eigenstates $\ket{\alpha\beta}$ and an interaction part $V$ that couples the channels. This results in a set of five coupled differential equations. With the Born--Oppenheimer potentials for the $S=0$ and $S=1$ states, the coupled equations can be solved numerically and from the calculated wavefunctions the position and widths of occurring Feshbach resonances can be predicted\cite{Mies}.

As shown in \cite{Mies}, the multichannel problem containing Feshbach resonances at well-separated magnetic field strengths can be reduced to an effective two-channel problem.

\subsection{Two-channel model}\label{sec:two_channel}
The two-channel model consists of an open channel and a single closed channel containing a molecular state, which couple to produce a resonance. In the center of mass reference frame, the coupled two-channel Hamiltonian is given by
\begin{equation}
	\hat{H}_{\mathrm{2b}}  =
	\begin{pmatrix}
		-\frac{\hbar^2}{2\mu_{2\mathrm{b}}}\frac{d^2}{dr^2}+v_{\mathrm{bg}}(r) & W(r)\\
		W(r) & -\frac{\hbar^2}{2\mu_{2\mathrm{b}}}\frac{d^2}{dr^2}+v_{\mathrm{c}}(r;B) 
	\end{pmatrix},
\end{equation}
where $\mu_{2\mathrm{b}}$ is the reduced mass of two $^{23}\mathrm{Na}$ atoms. The open channel potential $v_{\mathrm{bg}}$ serves as a reference, or background, and it is chosen to resemble the interactions of the scattering atoms in the $aa$ channel. This channel has mainly triplet character with $M_S=-1$. The potential should have a repulsive barrier at short distances and a characteristic van der Waals tail at large distances. The important properties of the open channel model potential are that its scattering length $a_{\mathrm{bg}}$ and the position of the last bound vibrational level agree with the values of the exact potential. We use the Lennard$-$Jones 6-10 potential with parameters for the triplet $^{23}\mathrm{Na}$ dimer \cite{BoGao_2003,Derevianko}
\begin{equation} \label{eq:LJ}
    v_{\mathrm{bg}}(r)=\frac{C_{10}}{r^{10}}-\frac{C_{6}}{r^6},
\end{equation}
with $C_{10}$ adjusted to support a single $s$-wave bound state with an $s$-wave scattering length of $63a_0$, where $a_0$ is the Bohr radius. The real sodium dimer does of course support many more bound states, but this simplified potential captures the essentials of the near-threshold $s$-wave scattering properties \cite{BoGao_2003}. The parameter settings and the potential properties are shown in \cref{table:sodium}. 

\begin{table}
	\centering
	\caption{LJ(6,10) model parameters for the two-channel, here $E_{\mathrm{vdW}}\approx1.18152\times10^{-8}$ Hartree.}
	\label{table:sodium}
	\begin{tabular*}{1.0\linewidth}{@{\extracolsep{\fill}}ccccc}
		\hline
		\hline
		$a_{\mathrm{bg}}(a_0)$&$C_6$(a.u.)&$C_{10}$(a.u.)&$r_{\mathrm{vdW}}(a_0)$&$E_{-1}/E_{\mathrm{vdw}}$\TBstrut\\
		\hline
		$63$&$1556$&$1.62005\times10^{9}$&$45$&$2.6$\TBstrut\\
		\hline
		\hline
	\end{tabular*}
\end{table}

The van der Waals length and energy, which characterize the range and scale of the potential, are defined via $C_6$ as 
\begin{align}\label{eq:vanderWaals}
r_{\mathrm{vdW}}&=\frac{1}{2}(2\mu_{2\mathrm{b}} C_6/\hbar^2)^{1/4},\\
E_{\mathrm{vdW}}&=\frac{\hbar^2}{2\mu_{2\mathrm{b}} r_{\mathrm{vdW}}^2},
\end{align}
and the mean scattering length \cite{Flambaum}, which sets the typical scale of the scattering length, is given by
\begin{equation}
\Bar{a}=\sqrt{2}(2\mu_{\mathrm{2b}}C_6/\hbar^2)^{1/4}\frac{\Gamma\left( 3/4 \right)}{\Gamma\left( 1/4 \right)}\approx 0.956r_{\mathrm{vdW}}.
\end{equation}
When  expressed in van der Waals units the potential (\ref{eq:LJ}) becomes
\begin{equation}
     v_{\mathrm{bg}}(r)=-\frac{16}{r^6}\left(1-\frac{C}{r^4}\right),   
\end{equation}
where the only non-universal constant $C=C_{10}/C_6r_{\mathrm{vdW}}^4$ determines $a_{\mathrm{bg}}$.

The uncoupled open channel wavefunction $\phi_{E}$ is defined as the energy-normalized regular solution to the Schr{\"o}dinger equation 
\begin{equation}
    \bigg(-\frac{\hbar^2}{2\mu_{\mathrm{2b}}}\frac{d^2}{dr^2}+v_{\mathrm{bg}}(r)\bigg)\phi_E(r)=E\phi_E(r),
\end{equation}
where $E$ is the scattering energy of the atoms. The asymptotic form of $\phi_{E}$ is given by
\begin{equation}
    \phi_E(r)\xrightarrow{ r \to \infty}\sqrt{\frac{2\mu_{\mathrm{2b}}}{\pi\hbar^2k}}\sin{(kr+\delta_{\mathrm{bg}})},
\end{equation}
where $k=\sqrt{2\mu_{\mathrm{2b}}E}/\hbar$ and $\delta_{\mathrm{bg}}$ is the background scattering phase shift. The background scattering length is given by
\begin{equation}
    a_{\mathrm{bg}}=\lim_{k \to 0} -\frac{\tan\delta_{\mathrm{bg}}}{k}.
\end{equation}
For all magnetic field strengths $B$, zero energy is defined as the asymptotic energy of the open channel.

For the closed channel it was shown in \cite{Mies} that the resonance is primarily caused by couplings to the channels $fh$ and $gg$, which, like $aa$, are both of triplet character, but with the important difference that $M_S=1$ (see \cref{table:Mchannels}). Thus the closed-channel potential differs from $v_{\mathrm{bg}}$ only by a magnetic-field dependent energy shift
\begin{equation}
    v_{\mathrm{c}}(r;B)=v_{\mathrm{bg}}(r)+E_{\mathrm{c}}(B),
\end{equation}
where we define the energy separation of the two channels as 
\begin{equation}
    E_{\mathrm{c}}(B)=E_{-1}+\delta\mu(B-B_{\mathrm{c}}).
\end{equation}
Here $E_{-1}$ is the binding energy of the only bound state of the background potential, and $\delta \mu$ is the difference in the magnetic moment between the separated atoms and the closed channel state. If the molecular states were of pure $M_S=\pm 1$ character one would have $\delta\mu=g_s \Delta M_s \mu_B=4\mu_B$, but due to mixing between different states, this is not exactly true. From \cite{Mies} we adopt the value $\delta \mu=3.8\mu_B$ in the vicinity of the resonances. 

The bare resonant state $\phi_{\mathrm{c}}$ is defined as the unit normalized solution to 
\begin{equation}
    \bigg(-\frac{\hbar^2}{2\mu_{\mathrm{2b}}}\frac{d^2}{dr^2}+v_{\mathrm{c}}(r;B)+E_{-1}\bigg)\phi_{\mathrm{c}}(r)=E_{\rm c}(B)\phi_{\mathrm{c}}(r).
\end{equation}
If there were no coupling between the channels, this state would be resonant with the open channel in the limit of zero collision energy when the magnetic field is tuned to $B_{\mathrm{c}}$.

The coupling between the channels occurs when the atoms are close, so the coupling term $W(r)$ should be a short-range function, otherwise, the exact form does not matter. We have used 
\begin{equation}\label{eq:interaction}
    W(r)=W_0\mathrm{e}^{-0.2r}.
\end{equation}
This coupling mixes states in the closed channel $\phi_{\mathrm{c}}$ with states in the open channel $\phi_E$, i.e., they become dressed. The dressed states are thus eigenstates to the Schr\"{o}dinger equation
\begin{equation}\label{eq:2b_schrodinger}
    (\hat{H}_{\mathrm{2b}}-E\hat{I})
	\begin{pmatrix}
		\phi_E(r) \\
		\phi_{\mathrm{c}}(r) 
	\end{pmatrix}=0,
\end{equation}
where $\hat{I}$ is the identity matrix. The coupling gives rise to a Feshbach resonance (i.e., a dressed resonant state) with an energy-dependent decay width \cite{Nygaard,Mies} given by
\begin{equation}\label{eq:decaywidth}
    \Gamma(E)=2\pi|\bracket{\phi_{\mathrm{c}}}{W|\phi_{E}}|^2.
\end{equation}

Due to the coupling, the magnetic field $B_0$ where the resonance occurs (i.e., the zero-energy crossing of the resonant state) is shifted somewhat from the resonant position of the bare resonant state $B_{\mathrm{c}}$. The two physical parameters of a Feshbach resonance are thus the resonance position $B_0$, and its measured width $\Delta B$, which are connected to the magnetic field strength dependent scattering length \cref{eq:Bdep_a}. 

The shift in the position of the resonance as it is dressed can be estimated \cite{Julienne_etAl,Mies_Raolt_Maurice} using multichannel quantum-defect theory
\begin{equation}\label{eq:B_shift}
    B_0-B_{\mathrm{c}}=\Delta B\frac{r_{\mathrm{bg}}(1-r_{\mathrm{bg}})}{1+(1-r_{\mathrm{bg}})^2},
\end{equation}
with $r_{\mathrm{bg}}=a_{\mathrm{bg}}/\bar{a}$.

The resonance width $\Delta B$ is the difference between the magnetic field at resonance $B_0$ and the magnetic field where $a=0$. It is connected to the decay width \cref{eq:decaywidth} through \cite{Nygaard,Mies}
\begin{equation}\label{eq:width}
    \Delta B=\lim_{k \rightarrow 0}\frac{\Gamma(E)}{2ka_{\mathrm{bg}}\delta\mu}.
\end{equation}

As a starting point we use the physical parameters $B_0$ and $\Delta B$ of the two Feshbach resonances in $^{23}\mathrm{Na}$ (see \cref{table:feshbach}) to extract the model parameter $W_0$ from \cref{eq:width} and $B_{\mathrm{c}}$ from \cref{eq:B_shift}.

\begin{figure}[ht]
\centering
    \includegraphics{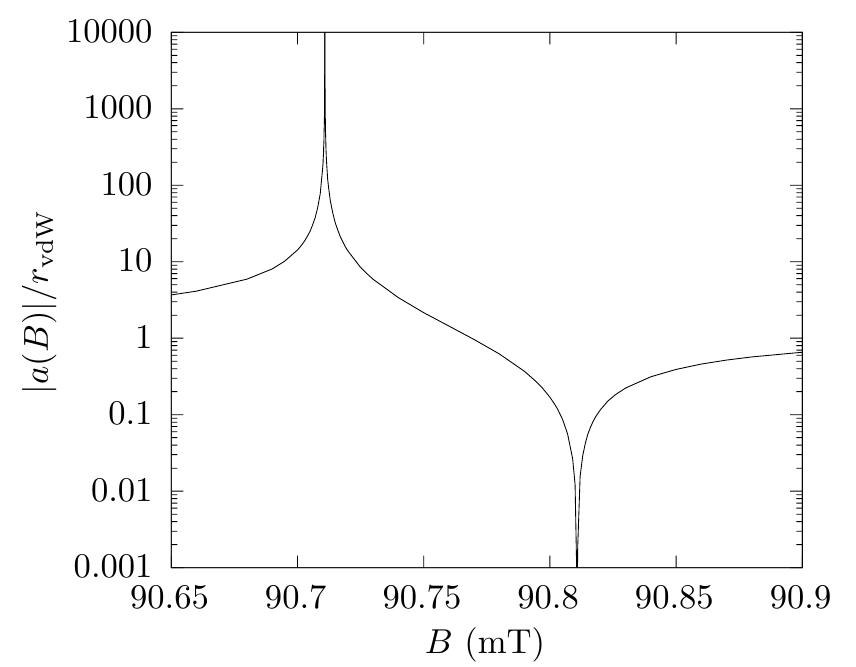}
    \caption{The magnitude of the scattering length $|a(B)|/r_{\mathrm{vdW}}$ plotted as a function of $B$ (mT) in the vicinity of the $s_{\mathrm{res}}=0.09$ resonance. The resonance position is slightly shifted from the input parameter $B_0=90.7$~mT to $B_0^*=90.71084$~mT.}
    \label{fig:scatl_sres_009}
\end{figure}

Assuming the value of $B_c$ from (\ref{eq:B_shift}), we then fine-tune the position of the resonance through an explicit calculation of the phase shift $\delta(E)$ in the dressed open channel, by solving \cref{eq:2b_schrodinger} using the R-matrix method outlined in \cite{Nygaard} with no boundary conditions at the R-matrix boundary $r_a$. The absolute value of the resulting $B$-dependent scattering length for the broader of the two resonances is shown in \cref{fig:scatl_sres_009}. Here, the resonance position is slightly shifted from the input parameter value $B_0=90.7$ mT to $B^*_0=90.71084$ mT but the width is almost identical. In \cref{table:feshbach} we list the fine-tuned magnetic fields $B_0^*$  giving zero-energy resonances.   

\subsection{Resonance strength}\label{sec:resonance_strength}
Feshbach resonances can be divided into two main types by introducing a dimensionless strength parameter $s_{\mathrm{res}}$ \cite{Chin_2010}. The resonance strength is defined by
\begin{equation}
s_{\mathrm{res}}=r_{\mathrm{bg}}\delta \mu \Delta B/\bar{E},
\end{equation}
where $\bar{E}=\hbar^2/2\mu_{2\mathrm{b}}\bar{a}^2$. Using $s_{\mathrm{res}}$ the intrinsic length $r^*$ \cite{Petrov_2004,Chin_2010} of the resonance can be defined as
\begin{equation}
\label{eq:rstar}
    r^*=\bar{a}/s_{\mathrm{res}}.
\end{equation}

Strong resonances are those with $s_{\mathrm{res}}>1$. The intrinsic length for these resonances is much smaller than the interaction length of the background potential, which for van der Waals interactions is given by $r_{\mathrm{vdW}}$. The strong resonances are also called open-channel dominated resonances because the atoms are more likely to reside in the entrance channel. This means that the spin of the dressed bound state (i.e., the Feshbach molecule) has more of the character of the open entrance channel over a large fraction of $\Delta B$ \cite{Chin_2010}. The strong resonances are often also broad in the sense that the resonance width $\Delta B$ is large.

Weak resonances are those with $s_{\mathrm{res}}\ll1$. Here $r^*$ is large compared to the background interaction length. These resonances are closed-channel dominated and they are usually narrow with regard to $\Delta B$.

\section{Three-Body scattering with a Feshbach resonance}
We here outline the theory of a four-channel model for the three-body scattering problem with a Feshbach resonance background. In all discussions below we assume that the total orbital angular momentum of the three-particle system is $J=0$, as is appropriate for Efimov states.

\subsection{Hyperspherical adiabatic representation}\label{sec:HAR}
We start by considering the one-channel problem for three particles in the absence of a Feshbach resonance. The Schr{\"o}dinger equation then takes the form
\begin{dmath}
\Bigg(-\frac{\hbar^2}{2}\sum_{i=1}^3 m_i^{-1}\Delta_{\vec{x}_i} + \sum_{i<j}^3 v(r_{ij})\Bigg)\Psi(\vec{x}_1,\vec{x}_2,\vec{x}_3)
=E\Psi(\vec{x}_1,\vec{x}_2,\vec{x}_3),
\label{eq:one-channel}
\end{dmath}
where $m_i$ are the particle masses, $v(r_{ij})$ is a two-body interaction potential and $r_{ij}=|\vec{x}_j-\vec{x}_i|$ is the distance between the particles $i$ and $j$.

We separate the center of mass motion from the internal motion of the particles by introducing mass-scaled Jacobi coordinates and then transform these internal systems to hyperspherical coordinates. With $M$ being the total mass,  the three-body reduced mass $\mu$ and a normalizing constant $d_k$ being defined as\footnote{ For identical particles of mass $m$, $\mu=m/\sqrt{3}$ and $d_k=\sqrt{2}/3^{1/4}$.}
\begin{equation}
    \begin{split}
        \mu^2&=\frac{m_im_jm_k}{M},\\
        d_k^2&=\frac{m_k(m_i+m_j)}{\mu M},
    \end{split}
\end{equation}
the mass scaled Jacobi coordinates and the center of mass coordinate are defined by
\begin{equation}
  \begin{split}
      \vec{r}_{k}&=d_k^{-1}(\vec{x}_j-\vec{x}_i),\\
      \vec{R}_{k}&=d_k(\vec{x}_k-(\vec{x}_j+\vec{x}_i)/2),\\
      \vec{X}_{\mathrm{c.m.}}&=\frac{1}{M} \sum_{i=1}^{3} m_{i} \vec{x}_{i}.
  \end{split}  
\end{equation}
Here, the indices $i,j,k$ are cyclic permutations of $(1,2,3)$.

Since the two-body interactions $v(r_{ij})$ are independent of $\vec{X}_{\mathrm{c.m.}}$, this transformation leads to a decoupling of the relative motion from the center of mass motion so that $\Psi(\vec{r}_{k},\vec{R}_{k},\vec{X}_{\mathrm{c.m.}})=\widetilde{\psi}(\vec{r}_{k},\vec{R}_{k})\varphi(\vec{X}_{\mathrm{c.m.}})$. The center of mass can thus be separated out. We continue with solving for the internal motion of the particles. 

The two Jacobi vectors are combined into a single position vector, whose components  define  a point in $\mathbb{R}^{6}$. The hyperspherical coordinates are the polar coordinates of this point, consisting of one hyperradius $\rho$ and five hyperangles $\Omega$. Three of these angles are external coordinates, which we choose to be the Euler angles $\alpha,\beta,\gamma$. They describe the orientation of the plane defined by the three particles, while the remaining two hyperangles $\theta$ and $\varphi_k$ describe the triangular shape of the system and the particle permutation at the vertices \cite{WANG20131}. Finally, the hyperradius describes the size of the system and is defined as $\rho=\sqrt{r^2_{k}+R^2_{k}}=3^{-1/4}\sqrt{r^2_{ij}+r^2_{jk}+r^2_{ki}}$. 

The rescaling $\widetilde{\psi}_n(\rho,\Omega)=\rho^{5/2}\psi_n(\rho,\Omega)$ removes first order derivatives with respect to $\rho$ from the hyperradial kinetic energy operator. The Schr\"{o}dinger equation for the internal motion is then
\begin{dmath}
\Bigg[-\frac{\hbar^2}{2\mu}\frac{\partial^2}{\partial \rho^2} + \frac{\hbar^2}{2\mu\rho^2}\Big(\Lambda^2+\frac{15}{4}\Big) +\sum_{i<j}^{3}V(r_{ij})\Bigg]\psi_n(\rho,\Omega)=E_n\psi_n(\rho,\Omega),
\label{eq:internal}
\end{dmath}
where $E_n$ is the energy eigenvalue and $\Lambda$ is the grand angular momentum. The dependence on the hyperangles is included in $\Lambda$, whose form depends on the definition of the hyperangles. 

For the internal hyperangles we use the mapping procedure described by Johnson \cite{Johnson1980} to define a modified set of the coordinates originally introduced by Smith and Whitten \cite{Smith_Whitten1968}. The two Jacobi vectors lie in the $\hat{x}\hat{y}$-plane, with the smaller moment of inertia along the $\hat{x}$-axis and the larger along the $\hat{y}$-axis. The $\hat{z}$-axis normal to this plane is positive in the direction of $\vec{A}=(\vec{r}_k \times \vec{R}_k)/2$ and $\hat{z}=\vec{A}/A$. The Cartesian components of the position vector are defined as
\begin{equation}\label{eq:smith_whitten}
    \begin{split}
        (\vec{r}_k)_x &= \rho \cos(\pi/4-\theta/2)\cos(\varphi_k/2),\\
        (\vec{r}_k)_y &= \rho \sin(\pi/4-\theta/2)\sin(\varphi_k/2),\\
        (\vec{r}_k)_z &= 0,\\
        (\vec{R}_k)_x &= -\rho \cos(\pi/4-\theta/2)\sin(\varphi_k/2),\\
        (\vec{R}_k)_y &= \rho \sin(\pi/4-\theta/2)\cos(\varphi_k/2),\\
        (\vec{R}_k)_z &= 0,
    \end{split}
\end{equation}
where $\theta \in [0,\pi/2]$ and $\varphi_k \in [0,4\pi)$. After requiring the wavefunction to be single-valued and imposing bosonic symmetry for identical particles, the range of $\varphi_k$ is reduced to $[0,\pi/3]$. For identical particles, kinematic rotations within the set of coordinates $(k,i,j)$ correspond to $\varphi_j=\varphi_i+4\pi/3$. Below we choose $k=3$ and suppress the indices on the hyperangular coordinate $\varphi_k$ in \cref{eq:smith_whitten}. Thus $\{ \vec{r}_1,\vec{R}_1  \}$ is obtained by $\varphi \rightarrow \varphi+4\pi/3$ and $\{ \vec{r}_2,\vec{R}_2  \}$ by $\varphi \rightarrow \varphi+8\pi/3$. The distances between the particles are then 
\begin{equation}\label{eq:distances}
    \begin{split}
        r_{12} &= \frac{d_k\rho}{\sqrt{2}}\big[1 + \sin\theta\cos\varphi\big]^{1/2},\\
        r_{23} &= \frac{d_k\rho}{\sqrt{2}}\big[1 + \sin\theta\cos(\varphi-2\pi/3)\big]^{1/2},\\
        r_{31} &= \frac{d_k\rho}{\sqrt{2}}\big[1 + \sin\theta\cos(\varphi+2\pi/3)\big]^{1/2}.
    \end{split}    
\end{equation} 

We now introduce the adiabatic representation by expanding the solutions $\psi_n(\rho,\Omega)$ into a complete set of orthonormal channel functions $\Phi_{\nu}(\rho;\Omega)$, with the radial wavefunctions $F_{\nu n}(\rho)$ as expansion coefficients. The channel functions are solutions to the adiabatic equation
\begin{equation}
    \Bigg[\frac{\hbar^2}{2\mu\rho^2}\Big(\Lambda^2+\frac{15}{4}\Big) +\sum_{i<j}^{3}V(r_{ij})\Bigg]\Phi_{\nu}(\rho;\Omega)=U_{\nu}(\rho)\Phi_{\nu}(\rho;\Omega),
    \label{eq:adiabatic}
\end{equation}
where the eigenvalues $U_{\nu}(\rho)$, here referred to as adiabatic potential curves, are obtained by solving \cref{eq:adiabatic} at a number of different hyperradii. Inserting the solution 
\begin{equation}
    \psi_n(\rho,\Omega)=\sum_{\nu=0}^{\infty}F_{\nu n}(\rho)\Phi_{\nu}(\rho;\Omega)
\end{equation}
into \cref{eq:internal}, projecting out $\Phi_{\mu}(\rho;\Omega)$, and integrating over the angles results in an infinite set of coupled ordinary differential equations
\begin{equation}
    \begin{split}\label{fullhamiltonian}
        \Bigg(-\frac{\hbar^2}{2 \mu}\frac{\partial^2}{ \partial \rho^2} + U_{\mu}(\rho) - \frac{\hbar^2}{2\mu}Q_{\mu\mu}(\rho) \Bigg)F_{n\mu}(\rho)&\nonumber\\ -\frac{\hbar^2}{2\mu}\Bigg(\sum_{\nu\neq\mu}2P_{\mu\nu}(\rho)\frac{\partial}{\partial\rho} + Q_{\mu\nu}(\rho) \Bigg)F_{n\nu}(\rho)& = E_nF_{n\mu}(\rho),
    \end{split}
\end{equation}
where $P_{\mu\nu}(\rho)$ and $Q_{\mu \nu}(\rho)$ are coupling matrix elements defined as 
\begin{equation}\label{eq:coupling}
\begin{split}
    P_{\mu\nu}(\rho) &= \int \,d\Omega \Phi_{\mu}^*(\rho;\Omega) \frac{\partial}{\partial\rho}\Phi_{\nu}(\rho;\Omega),\\
    Q_{\mu\nu}(\rho) &= \int \,d\Omega \Phi_{\mu}^*(\rho;\Omega) \frac{\partial^2}{\partial\rho^2}\Phi_{\nu}(\rho;\Omega),\\
    P^{2}_{\mu\nu}(\rho) &= -\int \,d\Omega \frac{\partial}{\partial\rho}\Phi_{\mu}^*(\rho;\Omega) \frac{\partial}{\partial\rho}\Phi_{\nu}(\rho;\Omega).
\end{split}
\end{equation}
The coupling matrix $\bar{P}$ is antisymmetric and therefore $P_{\nu\nu}=0$. The coupling matrices is \cref{eq:coupling} are related through
\begin{equation}
\bar{Q}=\frac{d\bar{P}}{d\rho}+\bar{P}^2,
\end{equation}
where the diagonal elements are $Q_{\nu\nu}=P^2_{\nu\nu}$.

The nonadiabatic coupling terms can be determined through numerical differentiation. However, this method is quite cumbersome. A more efficient method to calculate $P_{\mu\nu}$ uses the Hellmann--Feynman theorem \cite{Hellmann1933,Feynman}. Through an ingenious argument devised by Wang \cite{JiaWang}, this method can be extended to also calculate $P_{\mu\nu}^2$.

The adiabatic potentials $U_{\nu}$ contain most of the three-body physics and they can be viewed as three-body equivalents to the Born--Oppenheimer potentials in the two-body problem. The two-body recombination channels are associated with $U_{\nu}$ asymptotically approaching the energy of a two-body bound state, whereas for three-body continuum channels   
\begin{equation}\label{eq:continuum_channels}
  U_{\nu}(\rho)\xrightarrow{ \rho \to \infty}\hbar^2\frac{n(n+4)+15/4}{2\mu \rho^2}, 
\end{equation}
where $n=0,4,6,\ldots$ are eigenvalues of the asymptotically non-interacting problem, whose solutions can be expressed in terms of Gegenbauer polynomials. \cite{Blume}

We follow the common convention and define the three-body effective potential curves as 
\begin{equation}\label{eq:effective_potential}
    W_{\nu}(\rho)=U_{\nu}(\rho)-\frac{\hbar^2}{2\mu}Q_{\nu\nu}(\rho).
\end{equation}
In the asymptotic region $\rho\gg|a|$ the nonadiabatic couplings $Q_{\nu\nu}$ vanish and $W_{\nu}\rightarrow U_{\nu}$. 

At resonance $a\rightarrow\pm\infty$, one of the three-body effective potentials becomes attractive and takes the form
\begin{equation}\label{eq:efimov_attraction}
    W_{\nu}(\rho)=-\hbar^2\frac{|s_0|^2+1/4}{2\mu\rho^2},
\end{equation}
where $s_0\approx\pm1.00625i$. It is this attractive three-body potential that is responsible for the Efimov effect.

The bosonic symmetry is imposed through boundary conditions. From \cref{eq:distances} we find that the operation $\varphi\rightarrow -\varphi$ corresponds to $\{r_{12}\rightarrow r_{12}, r_{23}\leftrightarrow r_{31}\}$, while $\varphi-\pi/3\rightarrow -(\varphi -\pi/3)$ corresponds to $\{r_{31}\rightarrow r_{31}, r_{12}\leftrightarrow r_{23}\}$. (These  symmetries will be discussed in more detail below.) Bosonic identical-particle symmetry ensures that the wavefunction is unchanged under these operations. As a consequence, the derivatives at the reflection points have to vanish. That is, we solve \cref{eq:adiabatic} using the following boundary conditions 
\begin{equation}\label{eq:boundary_one_channelphi}   
    \frac{\partial}{\partial \varphi}\Phi_{\nu}(\rho;\theta,0) = \frac{\partial}{\partial \varphi}\Phi_{\nu}(\rho;\theta,\pi/3)= 0.
\end{equation}

The boundary conditions in $\theta$ are derived from the requirement that the solutions smoothly connect to the solutions for a vanishing potential (i.e., the hyperangular harmonics). In this limit, corresponding to the $\rho\rightarrow\infty$ limit of the adiabatic states, the hyperangular equation (\ref{eq:adiabatic}) becomes separable,
\begin{equation}
    -\Bigg(\frac{4}{\sin{2\theta}}\frac{\partial}{\partial\theta}\sin{2\theta}\frac{\partial}{\partial\theta}+\frac{4}{\sin^2{\theta}}\frac{\partial^2}{\partial\varphi^2}\Bigg) \Phi_{\nu}(\theta,\varphi)=\lambda_{\nu}\Phi_{\nu}(\theta,\varphi).
    \label{eq:adiabatic2}
\end{equation}
where $\lambda_{\nu}(\rho)\xrightarrow{\rho\rightarrow\infty}n(n+4)$. The solutions to \cref{eq:adiabatic2} with the $\varphi$-boundary conditions above have the form $\Phi_{\nu}(\theta,\varphi)=g_{l,m}(\theta)\cos{m \varphi}$ (where $m=0,3,6,\dots$ and $n=2(l+m)$), and
\begin{equation}
    \Bigg(-\frac{1}{\sin{2\theta}}\frac{\partial}{\partial\theta}\sin{2\theta}\frac{\partial}{\partial\theta}+\frac{m^2}{\sin^2{\theta}}\Bigg) g_{l,m}(\theta)=\frac{\lambda_{l,m}}{4}g_{l,m}(\theta).
    \label{eq:adiabatic3}
\end{equation}
 Regular solutions to this equation go as $\theta^m$ as $\theta\rightarrow0$ and approach a constant value as $\theta\rightarrow\pi/2$. Thus the boundary conditions in $\theta$ can be written as 
\begin{equation}\label{eq:boundary_one_channeltheta}
    \frac{\partial}{\partial \theta}\Phi_{\nu}(\rho;0,\varphi) = \frac{\partial}{\partial \theta}\Phi_{\nu}(\rho;\pi/2,\varphi) = 0.
\end{equation}

\subsection{Three-body physics in a four-state model}

We continue by considering the adiabatic part of the problem for three identical bosons, interacting via pair-wise interactions in the presence of a Feshbach resonance. For modeling the Feshbach resonance, we use the two-channel model described in \cref{sec:two_channel}. Each of the three atoms can then be in either one of two spin states: the open channel $\ket{o}$ or the closed molecular channel $\ket{c}$. For three atoms there are therefore in principle eight possible states. However, the atoms always get transferred from the open to the closed state in pairs, which means there has to be an even number of atoms in the closed state. So, for a three-body state, either all particles are in the open channel, or two particles are in the closed channel and one is in the open channel. The latter can be arranged in three different ways and the four three-body states are thus: $\ket{ooo}$, $\ket{coc}$, $\ket{occ}$, and $\ket{cco}$. Since we are including spin states we must now consider the symmetry requirements for the wavefunction to determine its boundary conditions. It is therefore necessary to take a short detour into the theory of permutation of particles.    

\subsubsection{Permutation of particles}
For a system of identical bosons, the total state must be invariant under an arbitrary permutation of particles. A system of $N$ particles has $N!$ permutations. The set of $N!$ permutation operators associated with this system forms the symmetric group $S_N$. 

Any permutation can be expressed as a product of transpositions\footnote{A transposition is a permutation that exchanges two particles.} and all transpositions are Hermitian and unitary. The transposition operator $(ij)$ exchanges the two particles in position $i$ and $j$. Since any transposition is unitary and any permutation is a product of transpositions, all permutations are also unitary. Not all permutations are Hermitian because the Hermitian conjugate of a product of operators changes the order of the operators and the transpositions do not necessarily commute. However, the Hermitian conjugate of a permutation is always its own inverse, as required by unitarity. Moreover, the decomposition of a permutation into a product of transpositions is not always unique. It will, however, always take either an even or an odd number of transpositions to write a specific permutation. A permutation is thus said to be even if built by an even number of transpositions, or odd if built by an odd number of transpositions. Within $S_N$, the number of even permutations equals the number of odd permutations.

It is possible to find states which are eigenstates of all elements in the permutation group. However, since the operators do not in general commute, this set cannot be complete, meaning that a general state cannot be expanded in this basis. Those that can are the states fulfilling bosonic or fermionic symmetry. The $N$ particle permutation operators are written $\hat{P}_{\alpha}=\hat{P}_{\alpha(1),\alpha(2)\ldots}$, where $\alpha$ is some permutation of the numbers $[1,2, \ldots, N]$.  The action of the permutation operator on an $N$-particle state is 
\begin{multline}
\hat{P}_{\alpha}\ket{a}_{(1)}\otimes \ket{b}_{(2)} \otimes \ket{c}_{(3)}\dots\\
\ket{a}_{(\alpha(1))}\otimes \ket{b}_{(\alpha(2))} \otimes \ket{c}_{(\alpha(3))}\dots,
\end{multline}
that is, $\hat{P}_{\alpha}$ takes the properties of particle 1 and gives them to particle $\alpha(1)$ and so on.
A completely symmetric state $\ket{\Psi_{\mathrm{S}}}$ is invariant under the action of all elements in the permutation group
\begin{equation}
\hat{P}_{\alpha}\ket{\Psi_{\mathrm{S}}}=\ket{\Psi_{\mathrm{S}}} \quad \forall \, \alpha.
 \end{equation}
These states are thus eigenstates to all operators in the group with unit eigenvalue. 
So in the total Hilbert space of $N$ particles, we can thus identify a subspace of symmetric states. It is possible to construct projectors onto such a symmetric subspace. With the permutation operators, we can build totally symmetric states using the symmetrizer
\begin{equation}
    \hat{S}=\frac{1}{N!}\sum_{\alpha}\hat{P}_{\alpha},
\end{equation}
where $\hat{S}=\hat{S}^{\dagger}$ and $\hat{S}^2=\hat{S}$. Bosonic symmetry implies that our states satisfy 
\begin{equation}
    \hat{S}\ket{\Psi}=\ket{\Psi}.
\end{equation}

\subsubsection{Permutation of three-body states}
We now consider permutations of three particle states. There are six elements in the symmetric group $S_3$, the identity, two cyclic permutations, and three transpositions. A three-particle state is represented by 
\begin{equation}
    \ket{a}_{(1)}\otimes \ket{b}_{(2)} \otimes \ket{c}_{(3)}\equiv\ket{abc},
\end{equation}
where the position in the ket refers to the particle number, i.e., particle $1$ has properties $a$, particle $2$ has properties $b$, and so on. Permutations of the particles in this state are performed by the permutation operator $\hat{P}_{ijk}$, which assigns the properties held by particle $1$ to particle $i$, $2 \rightarrow j$ and $3 \rightarrow k$. For example
\begin{equation}
    \hat{P}_{312}\ket{abc}=\ket{bca}.
\end{equation}
The permutation operators, their decomposition into products of transpositions, and their actions on kets and wavefunctions are listed in \cref{table:operators}. Henceforth, the operators $\hat{C}_n$ ($n=0-5$), defined in \cref{table:operators} are used for the different permutations. Since these operators form a group, they all have inverses. The identity and the transpositions are their own inverses, i.e., for $i=0,\ldots,3$, $\hat{C}_i^{-1}=\hat{C}_i$. For the cyclic permutations the inverses are $\hat{C}_4^{-1}=\hat{C}_5$ and $\hat{C}_5^{-1}=\hat{C}_4$. The Hermitian conjugates follow the same pattern, ensuring unitarity.

The Hermitian conjugates act on bra vectors from the right.  Thus for wavefunctions $\psi(\vec{x}_1,\vec{x}_2,\vec{x}_3)=\braket{\vec{x}_1 \vec{x}_2 \vec{x}_3|abc}$, where $\vec{x}_i$ is the position vector of particle $i$, the action of the cyclic permutation $\hat{C}_4$ is
\begin{dmath}
    \hat{C}_4\psi(\vec{x}_1,\vec{x}_2,\vec{x}_3)= \braket{\vec{x}_1 \vec{x}_2 \vec{x}_3|\hat{C}_4|abc}=\braket{\vec{x}_1 \vec{x}_2 \vec{x}_3|\hat{C}_5^{\dagger}|abc}=\psi(\vec{x}_3,\vec{x}_1,\vec{x}_2).
\end{dmath}
Thus, it is important to recognize that while the operators $\hat{C}_i$ act on the states, the operators $\hat{C}_i^\dagger$ act on the coordinates, and that for $i=4,5$ these are not the same.

{\setlength{\extrarowheight}{5pt}%
\begin{table}[ht]
\centering
\caption{The table summarizes the elements of $S_3$ and their respective action on kets and wavefunctions. Note that the two cyclic permutations $\hat{C}_4$ and $\hat{C}_5$ are each others Hermitian conjugates, i.e., $\hat{C}_4^{\dagger}=\hat{C}_5$ and $\hat{C}_5^{\dagger}=\hat{C}_4$.}
\begin{tabular*}{1.0\linewidth}{@{\extracolsep{\fill}}ll cc}
\hline
\hline
& & \multicolumn{2}{c}{Action on} \Tstrut\\
\cmidrule(lr){3-4}
Operator & Decomposition & $\ket{abc}$ & $\psi(\vec{x}_1,\vec{x}_2,\vec{x}_3)$ \Bstrut \\ 
\hline
$\hat{C}_0\equiv\hat{P}_{123}$ & $\mathbb{1}$ & $\ket{abc}$ & $\psi(\vec{x}_1,\vec{x}_2,\vec{x}_3)$ \Tstrut\\
$\hat{C}_1\equiv\hat{P}_{213}$ & $(12)$ & $\ket{bac}$ & $\psi(\vec{x}_2,\vec{x}_1,\vec{x}_3)$ \\
$\hat{C}_2\equiv\hat{P}_{132}$ & $(23)$ & $\ket{acb}$ & $\psi(\vec{x}_1,\vec{x}_3,\vec{x}_2)$\\
$\hat{C}_3\equiv\hat{P}_{321}$ & $(13)$ & $\ket{cba}$ & $\psi(\vec{x}_3,\vec{x}_2,\vec{x}_1)$ \\
$\hat{C}_4\equiv\hat{P}_{312}$ & $(23)(12)$\footnote{This decomposition is not unique. The other possible two-product decompositions being $(13)(23)$ and $(12)(13)$.} & $\ket{bca}$ & $\psi(\vec{x}_3,\vec{x}_1,\vec{x}_2)$ \\
$\hat{C}_5\equiv\hat{P}_{231}$ & $(12)(23)$\footnote{$(23)(13)$ and $(13)(12)$ are equivalent descriptions.} & $\ket{cab}$ & $\psi(\vec{x}_2,\vec{x}_3,\vec{x}_1)$\Bstrut \\
\hline
\hline
\end{tabular*}
\label{table:operators}
\end{table}}

\subsubsection{Symmetry transformations of Smith--Whitten coordinates}
For modeling the three-particle system it is advantageous to use the Smith--Whitten coordinates presented in \cref{eq:smith_whitten} since symmetry operations only affect the hyperangle $\varphi$. We shall now  discuss how inversions and permutations are carried out using these coordinates.

We have already in section \cref{sec:HAR} mentioned how transformations between different sets of Jacobi coordinates are performed. Here we select the coordinate system $k=3$ and investigate the effects of particle permutations.

Inversion symmetries, i.e., $\hat{\I}\{ \vec{r}_i,\vec{R}_i  \} \rightarrow \{ -\vec{r}_i,-\vec{R}_i  \}$, can be obtained in two ways, either by a transformation of the hyperangle $\varphi \rightarrow \varphi+2\pi$ or by a rotation of the coordinate system by an angle $\pi$ about the body-fixed $\hat{z}$-axis, i.e., the axis perpendicular to the plane defined by the three particles. Note that the inversion operator acts on coordinates and therefore leaves kets unchanged.

Rotations are obtained through changes of the Euler angles $(\alpha,\beta,\gamma)$ and lead to new body-fixed coordinate systems \cite{VMK}. The rotations of interest here are $\hat{D}_x(\pi)$, $\hat{D}_y(\pi)$ and $\hat{D}_z(\pi)$. They result in an inversion of the two axes that are not the axis of rotation, for example $\hat{D}_x(\pi)\{ \hat{x},\hat{y},\hat{z} \} = \{ \hat{x},-\hat{y},-\hat{z} \}$. 

A transposition of two identical particles is equivalent to an inversion of the Jacobi coordinate joining the two particles, i.e., $\{ \vec{r}_i,\vec{R}_i  \} \rightarrow \{ -\vec{r}_i,\vec{R}_i  \}$. All permutations in $S_3$ can be obtained by transformations of $\varphi$ combined with rotations of the entire system. However, the latter does not have a practical effect on our calculations since the wavefunction only depends on distances between bosons. 

There is no unique set of combinations of $\varphi$ transformations and rotations corresponding to each permutation in $S_3$, however, there is only one choice that results in $\varphi \in [0,4\pi]$. The actions of the Hermitian conjugates of these permutations on the coordinates are listed in \cref{table:actions}. As an example, the effect of $\hat{C}_4^{\dagger}$ can be illustrated as 
\begin{gather}\label{eq:triangle}
\begin{split}
    &\Biggl\langle\raisebox{-1.5ex}{\foo{scale=0.8}}\Bigg|\hat{C}_4^{\dagger}=\Biggl\langle\raisebox{-1.5ex}{\foo{scale=0.8}}\Bigg|\Big((23)(12)\Big)^{\dagger}\\
        &=\Biggl\langle\raisebox{-1.5ex}{\foo{scale=0.8}}\Bigg|(12)^{\dagger}(23)^{\dagger}
        =\Biggl\langle\raisebox{-1.5ex}{\fooo{scale=0.8}}\Bigg|(23)^{\dagger}=\Biggl\langle\raisebox{-1.5ex}{\foooo{scale=0.8}}\Bigg|.
\end{split}
\end{gather}
Note here that the transposition ($ij$) exchanges the particles in \emph{position} ($ij$). Thus $\hat{C}_4^\dagger $ transforms the Jacobi coordinates between each pair of particles according to $\{\vec{r}_3,\vec{r}_1,\vec{r}_2\}\rightarrow\{\vec{r}_1,\vec{r}_2,\vec{r}_3\}$ and
\begin{align*}
\bra{\varphi}\hat{C}_4^{\dagger}&=\bra{\varphi}(12)^{\dagger}(23)^\dagger\\
&=\bra{4\pi-\varphi}\hat{D}_y(\pi)^{\dagger}(23)^\dagger\\ &=\bra{4\pi-(10\pi/3-\varphi)}\hat{D}_y(\pi)^{\dagger}\hat{D}_x(\pi)^{\dagger}\\ & =\bra{2\pi/3+\varphi}\hat{D}_z(\pi)^{\dagger}.
\end{align*}
The transformation $\varphi \rightarrow \varphi +2\pi/3$ alone results in the correct transformation of the distances \cref{eq:distances}. However, this transformation in $\varphi$ needs to be combined with the rotation $\hat{D}_z(\pi)$ of the coordinate system to result in the correct transformation of the Jacobi vectors \cref{eq:smith_whitten}. 
{\setlength{\extrarowheight}{5pt}%
\begin{table}[h!]
\centering
\setlength{\extrarowheight}{7pt}
\begin{tabular*}{1.0\linewidth}{@{\extracolsep{\fill}}lllrrr}
\hline
\hline
& $\varphi$ & $\hat{D}$ & $\vec{r}_{3}$ & $\vec{r}_{1}$ & $\vec{r}_{2}$ \\ [0.5ex] 
\hline
$\hat{C}_0^{\dagger}$ & $\varphi$ & & $\vec{r}_{3}$ & $\vec{r}_{1}$ & $\vec{r}_{2}$ \\ 
$\hat{\I}$ & $2\pi+\varphi$ & & $-\vec{r}_{3}$ & $-\vec{r}_{1}$ & $-\vec{r}_{2}$ \\
$\hat{C}_1^{\dagger}$& $4\pi-\varphi$ & $\hat{D}_y(\pi)^{\dagger}$ & $-\vec{r}_{3}$ & $-\vec{r}_{2}$ &$-\vec{r}_{1}$\\
$(\hat{\I}\hat{C}_1)^{\dagger}$& $2\pi-\varphi$ & $\hat{D}_y(\pi)^{\dagger}$ & $\vec{r}_{3}$ & $\vec{r}_{2}$ &$\vec{r}_{1}$\\
$\hat{C}_2^{\dagger}$&$\dfrac{10\pi}{3}-\varphi$ & $\hat{D}_x(\pi)^{\dagger}$ & $-\vec{r}_{2}$ & $-\vec{r}_{1}$ & $-\vec{r}_{3}$ \\
$(\hat{\I}\hat{C}_2)^{\dagger}$&$\dfrac{4\pi}{3}-\varphi$ & $\hat{D}_x(\pi)^{\dagger}$ & $\vec{r}_{2}$ & $\vec{r}_{1}$ & $\vec{r}_{3}$ \\
$\hat{C}_3^{\dagger}$ & $\dfrac{8\pi}{3}-\varphi$ & $\hat{D}_y(\pi)^{\dagger}$ & $-\vec{r}_{1}$ & $-\vec{r}_{3}$ & $-\vec{r}_{2}$ \\
$(\hat{\I}\hat{C}_3)^{\dagger}$ & $\dfrac{2\pi}{3}-\varphi$ & $\hat{D}_y(\pi)^{\dagger}$ & $\vec{r}_{1}$ & $\vec{r}_{3}$ & $\vec{r}_{2}$ \\
$\hat{C}_4^{\dagger}$ & $\varphi+\dfrac{2\pi}{3}$& $\hat{D}_z(\pi)^{\dagger}$ & $\vec{r}_{1}$ & $\vec{r}_{2}$ & $\vec{r}_{3}$\\
$(\hat{\I}\hat{C}_4)^{\dagger}$ & $\varphi+\dfrac{8\pi}{3}$& $\hat{D}_z(\pi)^{\dagger}$ & $-\vec{r}_{1}$ & $-\vec{r}_{2}$ & $-\vec{r}_{3}$\\
$\hat{C}_5^{\dagger}$ & $\varphi+\dfrac{4\pi}{3}$ & & $\vec{r}_{2}$ & $\vec{r}_{3}$ & $\vec{r}_{1}$\\
$(\hat{\I}\hat{C}_5)^{\dagger}$ & $\varphi+\dfrac{10\pi}{3}$ & & $-\vec{r}_{2}$ & $-\vec{r}_{3}$ & $-\vec{r}_{1}$ \\[1.5ex]
\midrule
\end{tabular*}
\caption{The permutation operators $\hat{C}_n^{\dagger}$ are listed together with their corresponding transformation of the interparticle coordinates.}
\label{table:actions}
\end{table}}

\subsubsection{Symmetrization of three-body states}
By working with symmetrized states, we can restrict the domain of $\varphi$. This is hugely advantageous in numerical computations.

\begin{figure}[ht]
    \centering
    \includegraphics{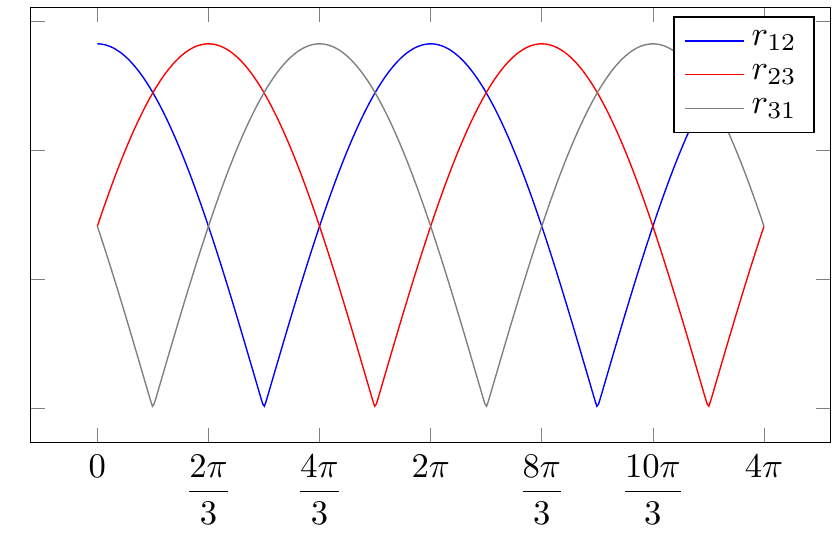}
    \caption{The interatomic distances defined in \cref{eq:distances} are here plotted as functions of $\varphi$ for $\theta=\pi/2$.}
    \label{fig:distances}
\end{figure}
\begin{figure}[H]
    \centering
    \includegraphics{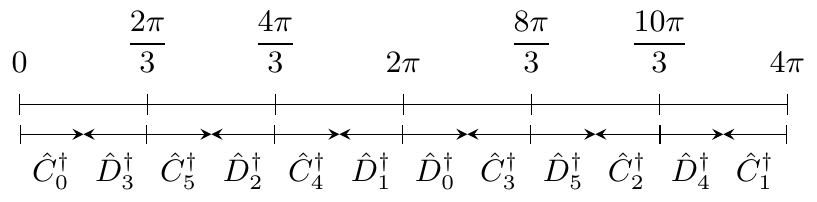}
    \caption{Illustration of the $\Tilde{\varphi}$ domain covered by applying $\hat{S}^\dagger(1+\hat{\I})$: $\varphi\in[0,\pi/3]$. The operators $\hat{D}_n^{\dagger}=(\hat{\I}\hat{C}_n)^{\dagger}$.}
    \label{fig:domain2}
\end{figure}

Using the labeling of the atoms given in \cref{eq:distances} and restricting the range of $\varphi$ to $[0,\pi/3]$, we effectively select the configuration where $r_{31} \leq r_{23} \leq r_{12}$ \footnote{Restricting the range $\varphi\in[2\pi,7\pi/3]$ selects the same pairs but with all the coordinates inverted.}, see \cref{fig:distances}. Clearly, this state is not symmetric. Applying the symmetrization operator $\hat{S}^\dagger$ to this state will not give a continuous coverage in $\varphi$, as is evident from the transformations in Table \ref{table:actions}. However, applying the combination $\hat{S}(1+\hat{\I})$ to $\ket{\varphi}$ defined on $[0,\pi/3]$ will result in a symmetrized state defined on the domain $\Tilde{\varphi}\in[0,4\pi]$, as is illustrated in Fig.\ \ref{fig:domain2}. The resulting state will thus be symmetrized not only with respect to particle permutations, but also parity. 

Solving the Schr{\"o}dinger equation on $\varphi \in [0,\pi/3]$ will result in a state
\begin{equation}\label{eq:tproductstate}
    \ket{\psi}=\ket{\phi}\otimes\ket{\xi},
\end{equation}
which consists of a spatial part, $\ket{\phi}$, and a spin part $\ket{\xi}$. The symmetrization operator then takes the form $\hat{S}=\hat{S}_{\phi}\otimes\hat{S}_{\xi}$. Because the inversion operator acts only on the spatial part,  the action  of $\hat{S}(1+\hat{\I})$ on \cref{eq:tproductstate} is
\begin{dmath}\label{eq:operator_on_tensorprod}
\bra{\varphi }\hat{S}(1+\hat{\I})\ket{\psi} =
\bra{\varphi}\hat{S}_{\phi}(1+\hat{\I})\ket{\phi}\otimes\hat{S}_{\xi}\ket{\xi}=\\
\frac{1}{N!}\sum_{i=0}^5\Big[\bra{\varphi}\hat{C}_{n}(1+\hat{\I})\ket{\phi}\otimes\hat{C}_{n}\ket{\xi}\Big].
\end{dmath}

We define the closed channel spatial parts by $\bracket{\varphi}{31}=\phi_{31}(\varphi)$ etc., and write the unsymmetrized state $\ket{\psi}$ as
 \begin{dmath}
        \bracket{\varphi}{\psi} = \phi_o(\varphi)\ket{ooo}+\phi_{31}(\varphi)\ket{coc}\\
        +\phi_{23}(\varphi)\ket{occ}+\phi_{12}(\varphi)\ket{cco}.
\end{dmath}
Here, $\phi_o$ is completely symmetric, but $\phi_{31}\neq\phi_{23}$ and $\phi_{23}\neq\phi_{12}$, since we have selected the configuration where the pair $\{3,1\}$ is most likely to form a molecule. 

From this solution on $\varphi \in [0,\pi/3]$, we can construct a general solution on $\Tilde{\varphi} \in [0,4\pi]$ using the pieces obtained by acting with $\hat{S}(1+\hat{\I})$. A symmetric wavefunction is thus obtained by
\begin{dmath}
    \bracket{\Tilde{\varphi}}{\psi_S}=\bra{\varphi}\hat{S}_{\phi}(1+\hat{\I})\ket{\phi}\otimes\hat{S}_{\xi}\ket{\xi}=\\
    \braket{\varphi|\hat{S}(1+\hat{\I})|\phi_o}\otimes\hat{S}_{\xi}\ket{ooo}+\\
    \braket{\varphi|\hat{S}_{\phi}(1+\hat{\I})|31}\otimes\hat{S}_{\xi}\ket{coc}+\\
    \braket{\varphi|\hat{S}_{\phi}(1+\hat{\I})|23}\otimes\hat{S}_{\xi}\ket{occ}+\\
    \braket{\varphi|\hat{S}_{\phi}(1+\hat{\I})|12}\otimes\hat{S}_{\xi}\ket{cco}.
\end{dmath}
For the spin parts in this equation, the non-conjugated operators are used. However, when working on the spatial parts we apply the transformation to $\varphi$, so here the corresponding conjugates are used. 

\begin{figure*}[ht]
    \centering
    \includegraphics{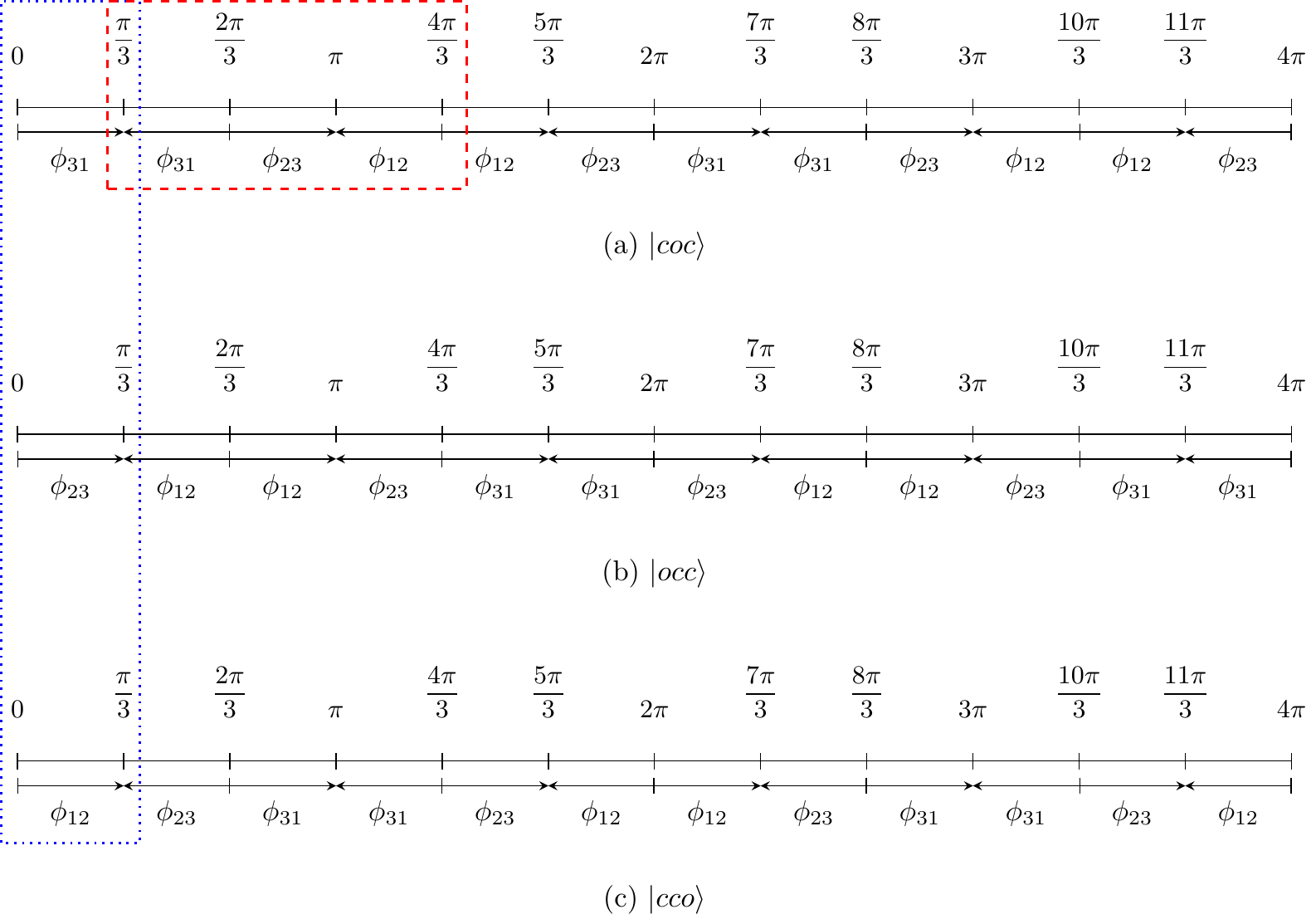}
    \caption{Illustration of the closed channel part of \cref{eq:piecewise}. The blue box encloses the region where $\varphi=\Tilde{\varphi}\in [0,\frac{\pi}{3}]$, with the closed channel part of the wave function then being $\phi_{31}\ket{coc}+\phi_{23}\ket{occ}+\phi_{12}\ket{cco}$. The red box in (a) illustrates how the functions evolve as $\Tilde{\varphi}$ goes from $\pi/3\to4\pi/3$.}
    \label{fig:equation}
\end{figure*}

We start by letting the operators act on each spin ket. For the three different closed channel parts, we then get 
\begin{equation}
    \begin{split}
    &\braket{\varphi|\hat{S}_{\phi}(1+\hat{\I})|31}\otimes\hat{S}_{\xi}\ket{coc}=\\
    \frac{1}{N!}\Big[&\braket{\varphi|(\hat{C}_0+\hat{C}_3)(1+\hat{\I})|31}\otimes\ket{coc}+\\
    &\braket{\varphi|(\hat{C}_1+\hat{C}_4)(1+\hat{\I})|31}\otimes\ket{occ}+\\
    &\braket{\varphi|(\hat{C}_2+\hat{C}_5)(1+\hat{\I})|31}\otimes\ket{cco}\Big],
    \end{split}
\end{equation}
\begin{equation}
    \begin{split}
        &\braket{\varphi|\hat{S}_{\phi}(1+\hat{\I})[|23}\otimes\hat{S}_{\xi}\ket{occ}]=\\
        \frac{1}{N!}\Big[&\braket{\varphi|(\hat{C}_0+\hat{C}_2)(1+\hat{\I})|23}\otimes\ket{occ}+\\
        &\braket{\varphi|(\hat{C}_3+\hat{C}_4)(1+\hat{\I})|23}\otimes\ket{cco}+\\
        &\braket{\varphi|(\hat{C}_1+\hat{C}_5)(1+\hat{\I})|23}\otimes\ket{coc}\Big]
    \end{split}
\end{equation}
and 
\begin{equation}
    \begin{split}
    &\braket{\varphi|\hat{S}_{\phi}(1+\hat{\I})[|12}\otimes\hat{S}_{\xi}\ket{cco}]=\\
    \frac{1}{N!}\Big[&\braket{\varphi|(\hat{C}_0+\hat{C}_1)(1+\hat{\I})|12}\otimes\ket{cco}+\\
    &\braket{\varphi|(\hat{C}_3+\hat{C}_5)(1+\hat{\I})|12}\otimes\ket{occ}+\\
    &\braket{\varphi|(\hat{C}_2+\hat{C}_4)(1+\hat{\I})|12}\otimes\ket{coc}\Big].
    \end{split}
\end{equation}
Combining the different closed channels and rewriting in terms of the conjugates gives
\begin{widetext} 
\begin{equation}
    \begin{split}
    &\bracket{\Tilde{\varphi}}{\psi_S}=\frac{1}{N!}\bigg\{\bracket{\varphi}{\phi_o}\otimes\ket{ooo}+\\
    \Big[&\braket{\varphi|(\hat{C}^{\dagger}_2+\hat{C}^{\dagger}_5)(1+\hat{\I}^{\dagger})|12}+\braket{\varphi|(\hat{C}^{\dagger}_1+\hat{C}^{\dagger}_4)(1+\hat{\I}^{\dagger})|23}+
    \braket{\varphi|(\hat{C}^{\dagger}_0+\hat{C}^{\dagger}_3)(1+\hat{\I}^{\dagger})|31}\Big]\otimes\ket{coc}+\\
    \Big[&\braket{\varphi|(\hat{C}^{\dagger}_3+\hat{C}^{\dagger}_4)(1+\hat{\I}^{\dagger})|12}+\braket{\varphi|(\hat{C}^{\dagger}_0+\hat{C}^{\dagger}_2)(1+\hat{\I}^{\dagger})|23}+
    \braket{\varphi|(\hat{C}^{\dagger}_1+\hat{C}^{\dagger}_5)(1+\hat{\I}^{\dagger})|31}\Big]\otimes\ket{occ}+\\
    \Big[&\braket{\varphi|(\hat{C}^{\dagger}_0+\hat{C}^{\dagger}_1)(1+\hat{\I}^{\dagger})|12}+\braket{\varphi|(\hat{C}^{\dagger}_3+\hat{C}^{\dagger}_5)(1+\hat{\I}^{\dagger})|23}+
    \braket{\varphi|(\hat{C}^{\dagger}_2+\hat{C}^{\dagger}_4)(1+\hat{\I}^{\dagger})|31}\Big]\otimes\ket{cco}\bigg\},
    \end{split}
\end{equation}
which results in
\begin{equation}\label{eq:piecewise}
\begin{split}
&\bracket{\Tilde{\varphi}}{\psi_S}=\frac{1}{N!}\bigg\{\bracket{\varphi}{\phi_o}\otimes\ket{ooo}+\\
\bigg[&\Big(\bra{\frac{10\pi}{3}-{\varphi}}+\bra{\frac{4\pi}{3}-{\varphi}}+\bra{{\varphi}+\frac{4\pi}{3}}+\bra{{\varphi}+\frac{10\pi}{3}}\Big)\ket{12}+\Big(\bra{4\pi-{\varphi}}+\bra{2\pi-{\varphi}}+\bra{{\varphi}+\frac{2\pi}{3}}+\bra{{\varphi}+\frac{8\pi}{3}}\Big)\ket{23}+\\
&\Big(\bra{{\varphi}}+\bra{{\varphi}+2\pi}+\bra{\frac{8\pi}{3}-{\varphi}}+\bra{\frac{2\pi}{3}-{\varphi}}\Big)\ket{31}
\bigg]\otimes\ket{coc}+\\
\bigg[&\Big(\bra{\frac{8\pi}{3}-{\varphi}}+\bra{\frac{2\pi}{3}-{\varphi}}+\bra{{\varphi}+\frac{2\pi}{3}}+\bra{{\varphi}+\frac{8\pi}{3}}\Big)\ket{12}+\Big(\bra{{\varphi}}+\bra{{\varphi}+2\pi}+\bra{\frac{10\pi}{3}-{\varphi}}+\bra{\frac{4\pi}{3}-{\varphi}}\Big)\ket{23}+\\
&\Big(\bra{4\pi-{\varphi}}+\bra{2\pi-{\varphi}}+\bra{{\varphi}+\frac{4\pi}{3}}+\bra{{\varphi}+\frac{10\pi}{3}}\Big)\ket{31}
\bigg]\otimes\ket{occ}+\\
\bigg[&\Big(\bra{{\varphi}}+\bra{{\varphi}+2\pi}+\bra{4\pi-{\varphi}}+\bra{2\pi-{\varphi}}\Big)\ket{12}+\Big(\bra{\frac{8\pi}{3}-{\varphi}}+\bra{\frac{2\pi}{3}-{\varphi}}+\bra{{\varphi}+\frac{10\pi}{3}}+\bra{{\varphi}+\frac{4\pi}{3}}\Big)\ket{23}+\\
&\Big(\bra{\frac{10\pi}{3}-{\varphi}}+\bra{\frac{4\pi}{3}-{\varphi}}+\bra{{\varphi}+\frac{2\pi}{3}}+\bra{{\varphi}+\frac{8\pi}{3}}\Big)\ket{31}
\bigg]\otimes\ket{cco}\bigg\}
.
\end{split}
\end{equation}
\end{widetext}
A graphical representation of the closed channel part of \cref{eq:piecewise} is shown in \cref{fig:equation}. As mentioned previously, the magnitudes of the amplitudes are in the following order $\phi_{31}\geq\phi_{23}\geq\phi_{12}$. Here, the blue box encloses the region where $\varphi=\Tilde{\varphi}\in [0,\frac{\pi}{3}]$ and the closed channel part of the wavefunction is then $\phi_{31}\ket{coc}+\phi_{23}\ket{occ}+\phi_{12}\ket{cco}$. The red box in \cref{fig:equation}(a) illustrates how the functions evolve from $\phi_{31}\to\phi_{23}\to\phi_{12}$ as $\Tilde{\varphi}$ grows from $\pi/3\to4\pi/3$. This behavior can also be traced in \cref{fig:distances}, where the shortest atomic distance corresponds to the spin channel with the largest amplitude, i.e., $\phi_{31}$. By examining the distances in \cref{fig:distances} over the range $[\frac{\pi}{3},\frac{4\pi}{3}]$, we can infer that channel $\ket{coc}$ will have the amplitude $\phi_{31}$ as $\Tilde{\varphi}\in[\frac{\pi}{3},\frac{2\pi}{3}]$ and $\ket{cco}$ will have the amplitude $\phi_{31}$ as $\Tilde{\varphi}\in[\frac{2\pi}{3},\frac{4\pi}{3}]$. 
We use the graph of the $\ket{coc}$ channel in \cref{fig:equation}(a) and write the spatial part as the piecewise defined function
\begin{equation}
    \Phi_C(\Tilde{\varphi})\equiv
    \begin{cases}
    \phi_{31}(\Tilde{\varphi}), &\text{if } \Tilde{\varphi}\in[0,\frac{\pi}{3}],\\
    \phi_{31}(\frac{2\pi}{3}-\Tilde{\varphi}), &\text{if } \Tilde{\varphi}\in[\frac{\pi}{3},\frac{2\pi}{3}],\\
    \phi_{23}(\Tilde{\varphi}-\frac{2\pi}{3}), &\text{if } \Tilde{\varphi}\in[\frac{2\pi}{3},\pi],\\
    \phi_{12}(\frac{4\pi}{3}-\Tilde{\varphi}), &\text{if } \Tilde{\varphi}\in[\pi,\frac{4\pi}{3}],\\
    \phi_{12}(\Tilde{\varphi}-\frac{4\pi}{3}), &\text{if } \Tilde{\varphi}\in[\frac{4\pi}{3},\frac{5\pi}{3}],\\
    \phi_{23}(2\pi-\Tilde{\varphi}), &\text{if } \Tilde{\varphi}\in[\frac{5\pi}{3},2\pi].
    \end{cases}
\end{equation}
In this way, the spatial part is expressed in terms of our original channel functions $\phi_{12}(\varphi)$, $\phi_{23}(\varphi)$, and $\phi_{31}(\varphi)$, which were defined on $\varphi \in [0,\frac{\pi}{3}]$.

Using $\Phi_C$, the closed channel terms can be expressed more compactly using shifted versions of \cref{fig:equation}(a):
\begin{equation}
    \Phi_C(\Tilde{\varphi})\ket{coc}
  +  \Phi_C(\Tilde{\varphi}+4\pi/3)\ket{occ}+\Phi_C(\Tilde{\varphi}+2\pi/3)\ket{cco}.
\end{equation}
 
Finally, to determine the appropriate boundary conditions, we can look at the $\ket{coc}$ channel over the range $\Tilde{\varphi}\in[\frac{\pi}{3},\frac{4\pi}{3}]$, since this is the first $\pi$-long interval that contains all our sub-functions $\phi_{31}$, $\phi_{23}$ and $\phi_{12}$.

For $\phi_{o}$ the boundary conditions are the same as in the one-channel problem, see \cref{eq:boundary_one_channelphi}. However, for the closed channel subfunctions, the situation is different. At the endpoints of $\Tilde{\varphi}\in[\frac{\pi}{3},\frac{4\pi}{3}]$ we should have $\Phi_C'(\Tilde{\varphi})=0$ and $\Phi_C$ should be smooth at $\Tilde{\varphi}=\frac{2\pi}{3}$ and $\Tilde{\varphi}=\pi$. This equates to the following boundary conditions 
\begin{equation} \label{eq:bc0}
\text{at } \varphi=0:
    \begin{cases}
    \phi_{o}'(0)=0,\\
    \phi_{12}'(0)=0,\\
    \phi_{31}(0)=\phi_{23}(0),\\
    \phi_{31}'(0)=\phi_{23}'(0),\\
    \end{cases}
\end{equation}
\begin{equation} \label{eq:bcpi3}
\text{at } \varphi=\frac{\pi}{3}:
    \begin{cases}
    \phi_{o}'(\frac{\pi}{3})=0\\
    \phi_{31}'(\frac{\pi}{3})=0\\
    \phi_{23}(\frac{\pi}{3})=\phi_{12}(\frac{\pi}{3}),\\
    \phi_{23}'(\frac{\pi}{3})=\phi_{12}'(\frac{\pi}{3}).
    \end{cases}
\end{equation}

\subsection{The four-state Hamiltonian and numerical implementations}

When we include the three channels containing atoms in the closed channel, the dimension of our Hamiltonian matrix grows by a factor of four. Thus,
the four-state Hamiltonian is then given by
\begin{equation} \label{eq:H}
\mathrm{\hat{H}}=
\begin{pmatrix}
\hat{H}_{\mathrm{bg}} & W_{31}  & W_{23} & W_{12}\\
 W_{31}  & \hat{H}_{\mathrm{c}_{31}}&0&0\\
 W_{23}  &0& \hat{H}_{\mathrm{c}_{23}}&0\\
 W_{12}  &0& 0&\hat{H}_{\mathrm{c}_{12}}
\end{pmatrix}.
\end{equation}
Here the open-channel Hamiltonian is 
\begin{equation}
\hat{H}_{\mathrm{bg}}=-\frac{\hbar^2}{2\mu}\frac{\partial^2}{\partial\rho^2}+\hbar^2\frac{\Lambda^2+15/4}{2\mu\rho^2}+V_{\mathrm{bg}},
\end{equation}
where $V_{\mathrm{bg}}=v_{\mathrm{bg}}(r_{12})+v_{\mathrm{bg}}(r_{23})+v_{\mathrm{bg}}(r_{31})$ is the background potential and the closed-channel Hamiltonian is
\begin{equation}
    	\hat{H}_{\mathrm{c}_{ij}}=-\frac{\hbar^2}{2\mu}\frac{\partial^2}{\partial\rho^2}+\hbar^2\frac{\Lambda^2+15/4}{2\mu\rho^2}+v_{c}(r_{ij}),
\end{equation}
where $v_c(r_{ij};B)$ is the closed-channel potential for the pair $ij$. The function $W_{ij}$ couples the closed $ij$ channel to the open channel. Note that we only include pairwise interactions between the atoms, which is usually deemed sufficient for capturing the essentials of Efimov physics. A fully realistic representation of the three-atom system would not only require a far more elaborate potential than the model \cref{eq:LJ}, but also the inclusion of nonadditive short-range three-body forces \cite{Soldan2003}. Since our model potential does not have the true short-range form anyway, there is no reason to include this additional complexity. This also applies to the interaction between a pair of atoms bound in a closed channel and the third atom in the open channel. Any realistic representation of this atom-molecule interaction would not resemble the sum of two potentials of the form \cref{eq:LJ}.

In \cref{eq:H} the rows and columns correspond to the channels $\ket{ooo}$, $\ket{coc}$, $\ket{occ}$, and $\ket{cco}$, and similarly, its eigenfunctions take the form 
\begin{equation}
    \begin{pmatrix}
        \phi_{o}(\theta,\varphi)\ket{ooo}\\
        \phi_{31}(\theta,\varphi)\ket{coc}\\
        \phi_{23}(\theta,\varphi)\ket{occ}\\
        \phi_{12}(\theta,\varphi)\ket{cco}\\
    \end{pmatrix}.
\end{equation}

In our numerical implementation, we expand all channel functions $\phi_\alpha$ in the same set of basis functions. The implementation of the boundary conditions becomes particularly simple using a basis of B-splines $B_{i,k}$. Thus, the channel functions $\phi_\alpha$ are expanded as
\begin{equation}
    \label{eq:basis}   \phi_\alpha(\theta,\varphi)=\sum_{i=1}^{N_\varphi}\sum_{j=1}^{N_\theta}c_{ij}^\alpha B_{i,k}(\varphi)B_{j,k}(\theta),
\end{equation}
where $\alpha=o, 31, 23$ or $12$. The B-splines have the following useful properties
\begin{align*}
    &B_{1,k}(0)=1, \qquad &&B_{i>1,k}(0)=0, \\
   &B_{N_\varphi,k}(\pi/3)=1, \qquad &&B_{i<N_\varphi,k}(\pi/3)=0, \\
    &B_{1,k}'(0)=-B_{2,k}'(0),\\  
    &B_{N_\varphi,k}'(\pi/3)= -B_{N_\varphi-1,k}'(\pi/3). 
\end{align*}
The boundary conditions in \cref{eq:bc0,eq:bcpi3} can then be enforced by setting (for all $j$)
\begin{align*}
&c_{1,j}^o=c_{2,j}^o, &&c_{1,j}^{12}=c_{2,j}^{12},\\
&c_{1,j}^{31}=c_{1,j}^{23}, &&c_{2,j}^{31}=c_{2,j}^{23},\\
&c_{N_\varphi,j}^{12} = c_{N_\varphi,j}^{23}, &&c_{N_\varphi-1,j}^{12} = c_{N_\varphi-1,j}^{23},\\
&c_{N_\varphi,j}^o = c_{N_\varphi-1,j}^o, &&c_{N_\varphi,j}^{31} = c_{N_\varphi-1,j}^{31}.
\end{align*}
Thus the boundary conditions reduce the sum over basis functions in $\varphi$ from $N_\varphi$ terms to $N_\varphi-2$ terms, while building the boundary conditions into the basis functions. The price to pay is that when the Hamiltonian in \cref{eq:H} is expressed in this reduced basis it will not have a simple block-diagonal form anymore. As an example, the basis function with coefficient $c_{1,j}^{31}$ will overlap not only with functions that are multiplied by coefficients $c_{i>1,j}^{31}$ but also with functions having coefficients of type $c_{i,j}^{23}$.

\section{Results and Discussion}

The main topic in this section regards features discovered in the three-body adiabatic potentials that affect the position of the repulsive wall in the effective potential responsible for the Efimov effect. The appearance of this wall at a seemingly universal position $\rho\approx2r_{\mathrm{vdW}}$ in single-channel models utilizing a range of different two-body interactions \cite{Wang_2012_Origin} has been used to explain the experimentally observed universality of the position of the appearance of the first Efimov resonance from the three-body continuum. The universality of the scattering length where the Efimov resonance couples to the three-body continuum ($a_-$) was first thought to be exclusive for open-channel dominated resonances with $s_{\mathrm{res}}>1$. However, experiments have shown that the universality persists in an intermediate resonance regime where $s_{\mathrm{res}}\simeq 0.1$ \cite{ Roy2013}. An even weaker resonance, a heteronuclear resonance in $^7{\mathrm{Li}}-^{13  3}{\mathrm{Cs}}$ with $s_{\mathrm{res}}=0.05$ has shown a clear deviation from this universality, with a scattering length for a zero-energy Efimov state larger than predicted by van der Waals universality, suggesting a trend of increasing $a_-$ with decreasing $s_{\mathrm{res}}$ \cite{Johansen2017}. While calculations of $a_-$ are outside the scope of this paper, it is still interesting to examine the form of the Efimov potential, since the onset of the repulsive wall and the depth of the potential well correlate with this parameter. 

To place our work in the context of previous theoretical discoveries, 
we begin with a short review of how this universality manifests in single-channel models and discuss its physical underpinnings. We then present our findings within the multi-channel model and show how tuning the Feshbach resonance to the closed-channel dominated regime affects couplings of the adiabatic potentials, which subsequently leads to changes in the effective potentials causing the Efimov effect. We show that our discoveries are in line with the experimentally found trend as well as with recent theoretical predictions based on numerical findings with a different kind of two-channel Feshbach model \cite{Kraats_Kookelmans}.

\subsection{Prelude}\label{sec:Prelude}
The appearance of a universally positioned repulsive wall in the hyperradial three-body potentials, obtained using a single-channel model based on two-body interactions with attractive van der Waals tails, was discovered by Wang et al. \cite{Wang_2012_Origin}. They argued that the experimentally observed universality of $a_-$, i.e., the unexpected universality of the three-body parameter, is caused by this universal repulsion in the three-body effective potential. The repulsion, located in the intermediate hyperradial region at $\rho\approx2r_{\mathrm{vdW}}$, prevents three-body states from probing the short-range region, i.e., $\rho<r_{\mathrm{vdW}}$, where species-specific interactions would otherwise occur. This effectively makes any three-body observable insensitive to the details of the two-body interaction. The fact that the repulsion occurs at a specific distance points to a possible cause for the universality of the three-body parameter for atomic collisions governed by the van der Waals force. 

The universal three-body repulsion can be explained by an effect appearing at the two-body level, where there is a suppression of the probability of finding any pair of atoms at separations shorter than the characteristic length of the van der Waals attraction, i.e., at separations shorter than $r_{\mathrm{vdW}}$. The collective pair suppression at atomic separations close to $r_{\mathrm{vdW}}$ forces the state into an equilateral triangular shape as $\rho$ is decreased past a certain point. The mechanism by which this two-body suppression-driven deformation gives rise to a sudden nonadiabatic increase in kinetic energy was investigated in \cite{NaidonPascalUeda}. At the onset of deformation, the nonadiabatic diagonal term in \cref{eq:effective_potential} becomes large. This results in a repulsive wall in the effective three-body potential and, if the pair suppression is strong and universal with respect to $r_{\mathrm{vdW}}$, it will appear at a more or less fixed hyperradius when scaled by $r_{\mathrm{vdW}}$. 

\begin{figure}[ht]
    \centering
    \includegraphics{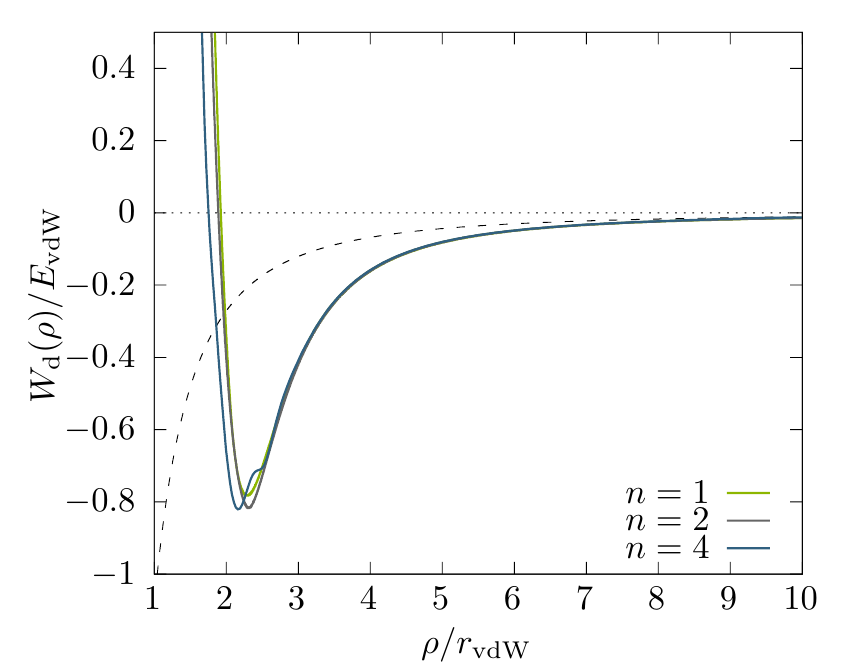}
    \caption{Efimovian effective potentials obtained from a single-channel calculation using a two-body Lennard$-$Jones 6-10 model with $n=1,2$ and 4 $s$-wave states and $a=\pm\infty$.}
    \label{fig:single_channel}
\end{figure}

In \cref{fig:single_channel} we show the three-body effective potentials $W_{\mathrm{d}}$ obtained from a single-channel model using the Lennard--Jones potential given in \cref{eq:LJ} and $C_{10}$ tuned to contain $n=1,2$ or 4 $s$-wave bound states, and $a=\pm\infty$. The potentials have been diabatized through an avoided crossing as in \cite{Wang_2012_Origin}. It should be noted that the potentials obtained here, using only a few $s$-wave bound states, are very similar to the ones in \cite{Wang_2012_Origin,NaidonPascalUeda}, both with regard to the position of the zero-energy crossing (i.e., the repulsive wall or barrier), the energy of the minimum, and the long-range attractive tail. An analysis of the pair amplitude suppression inside the relevant two-body potential well yields practically the same result for $n=1$ and $n=8$ $s$-wave bound states and confirms a strong two-body suppression for $r<r_{\mathrm{vdW}}$ with the potential containing only one bound state. We, therefore, conclude that at the two-body level, the pair suppression with one $s$-wave bound state is sufficient to (at least to a good approximation) retrieve the single-channel universal Efimov potential with van der Waals universality.

\subsection{Four-state model results}
The results presented here concern the three-body adiabatic potentials $U_\nu$ and the effective potentials $W_\nu$ obtained by diagonalizing the adiabatic four-state Hamiltonian in a basis of B-splines (typically 90 in both $\theta$ and $\varphi$, thus giving a $26896\times 26896$ Hamiltonian) at different hyperradii. We use a notational convention in which the sign of $\nu$ indicates the asymptotic configuration of the three atoms. A negative sign corresponds to an atom-dimer channel, a positive sign to a three-body continuum channel, and $\nu=0$ is the channel that converges to the universal Efimov attraction \cref{eq:efimov_attraction}.

The effective potentials $W_\nu$ represent an improved approximation compared to the adiabatic potentials $U_\nu$. For many adiabatic states the difference is relatively small but close to threshold the two levels of approximation show qualitatively different features. Indeed the van der Waals universality appears only in $W_0$. Still, it should be emphasized that the hyperangular wavefunction $\Phi_\nu(\rho;\Omega)$ is the same in both approximations, namely, the eigenfunction of the hyperangular Hamiltonian \cref{eq:adiabatic}. The difference being that while $U_\nu$ is the corresponding eigenvalue, the effective potential $W_\nu$ is the expectation value of the full Hamiltonian, including the hyperradial kinetic energy. Thus, properties associated with the hyperangular configuration as well as those connected to the four Feshbach channels remain unchanged. As we will see below, adding the correction term makes the energy landscape very complex. Therefore we have found it useful to first discuss the features associated with the adiabatic states, before moving on to the consequences this has for the effective potentials.

The Feshbach model was built to resemble two of the resonances found in $^{23}\mathrm{Na}$, occurring at about $90.7$ mT and $85.3$ mT. The physical properties of these and the parameters used in the two-body model have been well charactrized in \cite{Mies}, and are listed in the top two rows of \cref{table:feshbach}. The broader one of the resonances is of intermediate strength with $s_{\mathrm{res}}=0.09$, while the narrow one is weak with $s_{\mathrm{res}}=0.0002$. We refer to these as the intermediate and the narrow resonance. 

\subsubsection{Three-body adiabatic potentials}
\begin{figure}[t]
    \centering
    \includegraphics{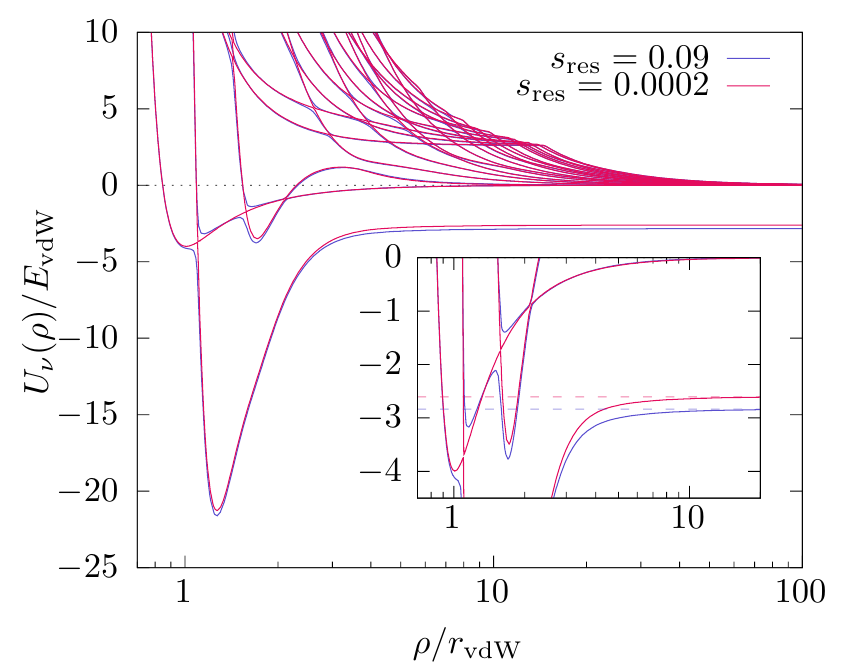}
    \caption{The 20 lowest three-body adiabatic potential curves $U_{\nu}$ for the intermediate ($s_{\mathrm{res}}=0.09$) and the narrow ($s_{\mathrm{res}}=0.0002$) resonance. The corresponding dressed state two-body energies are shown as dashed lines in the inset figure.}
    \label{fig:energymultiplot}
\end{figure}

In \cref{fig:energymultiplot} the 20 lowest lying three-body adiabatic potential curves $U_{\nu}$ for the intermediate (blue) and narrow (red) resonances are shown. For both potentials, the magnetic field was set to $B_0^*$, i.e., $a\rightarrow\pm\infty$. At a first glance the curves look similar. The lowest lying potential curves $U_{-1}$ can be seen to converge to the energy of the corresponding dressed dimer, which is $-2.8E_{\mathrm{vdW}}$ for the intermediate and $-2.6E_{\mathrm{vdW}}$ for the narrow resonance (indicated by the dashed horizontal lines in the zoomed-in inset figure). The lowest adiabatic potentials for the two resonances differ from each other more markedly in the region $\rho\sim r_{\mathrm{vdW}}$. Here, the potentials obtained with a narrow resonance background have three very sharp avoided crossings, whereas for the intermediate resonance the avoided crossings, except the one at approximately $2r_{\mathrm{rvdW}}$, are much softer.

To examine the Efimovian long-range features of the potentials it is more practical to rewrite the potentials as  
\begin{equation}\label{eq:lambda_nu}
    \lambda_{\nu}(\rho)=\frac{2\mu}{\hbar^2}\rho^2 U_{\nu}(\rho)+\frac{1}{4}.
\end{equation}
For resonant two-body interactions ($a=\pm\infty$), the channel that causes the Efimov effect can then be recognized as the one that converges to $-s_0^2\approx-1.0125$ in the so-called scale-free region $\rho\gg r_{\mathrm{c}}$ (for finite $a$, the region where $|a|\gg\rho\gg r_{\mathrm{c}}$). Here, $r_{\mathrm{c}}$ is the characteristic length of the two-body interaction, which for open channel dominated resonances with intrinsic length  $r^*<r_{\mathrm{vdW}}$ (see Eq.\ (\ref{eq:rstar})) is given by $r_{\mathrm{vdW}}$ and for closed channel dominated resonances with $r^*\gg r_{\mathrm{vdW}}$ is given by $r^*$. 

\begin{figure}[ht]
    \centering
    \includegraphics{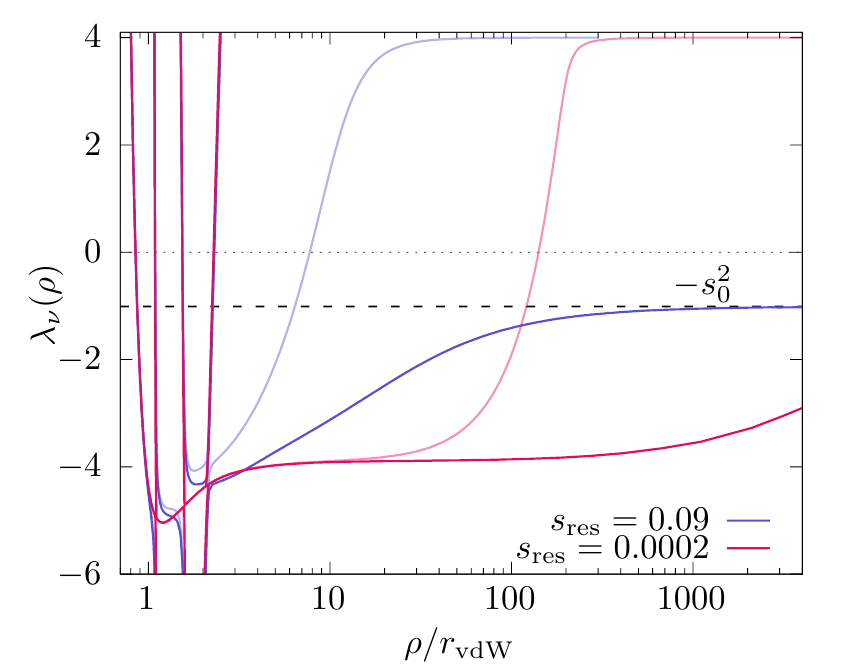}
    \caption{Three-body potentials $\lambda_{\nu}$ ($\nu=-1$ to $1$) versus $\rho/r_{\mathrm{vdW}}$ for the broad and narrow resonances with $a=\pm\infty$ (full-color curves) and $a=0$ (light curves).}
    \label{fig:lambda0}
\end{figure}
In \cref{fig:lambda0} we show the potentials $\lambda_{\nu}$ ($\nu=-1$ to $1$) for the intermediate and narrow resonance with resonant interactions, together with the corresponding curves for effectively non-interacting atoms ($a=0$) (light curves). Our calculations extend to $\rho=4000r_{\mathrm{vdW}}$, beyond this limit we were not able to obtain converged results. For resonant two-body interactions, the potential $\lambda_0$ for the intermediate resonance is clearly converging to the universal Efimov potential $-s_0^2$ as $\rho > 300 r_{\mathrm{vdW}}$. The intrinsic length of this resonance is $r^*\approx 10 r_{\mathrm{vdW}}$, so the Efimovian features of this potential appear for hyperradii that are about one order of magnitude larger than $r^*$. Since $r^*\approx 4000 r_{\mathrm{vdW}}$ for the narrow resonance, the scale-free region has not been reached and the potentials should therefore not exhibit the Efimovian long-range form. Our findings are thus in line with the predictions made by Petrov in \cite{Petrov_2004}. 

The behavior of $\lambda_0$ for $a=0$ at large $\rho$ is, for sufficiently large separations,  equivalent to that of three non-interacting atoms. The asymptotic form of a three-body continuum channel is $\lambda_{\nu}\rightarrow n(n+4)+4$ (see \cref{eq:lambda_nu,eq:continuum_channels}). Since $\lambda_0$ is converging to 4 in the long-range region, this channel corresponds to the lowest three-body continuum channel with the Gegenbauer eigenvalue $n=0$. Remarkably, for the narrow resonance with vanishing scattering length the atoms interact all the way out to $\rho\approx100r_{\mathrm{vdW}}$.

\begin{figure}[ht]
    \centering
    \includegraphics{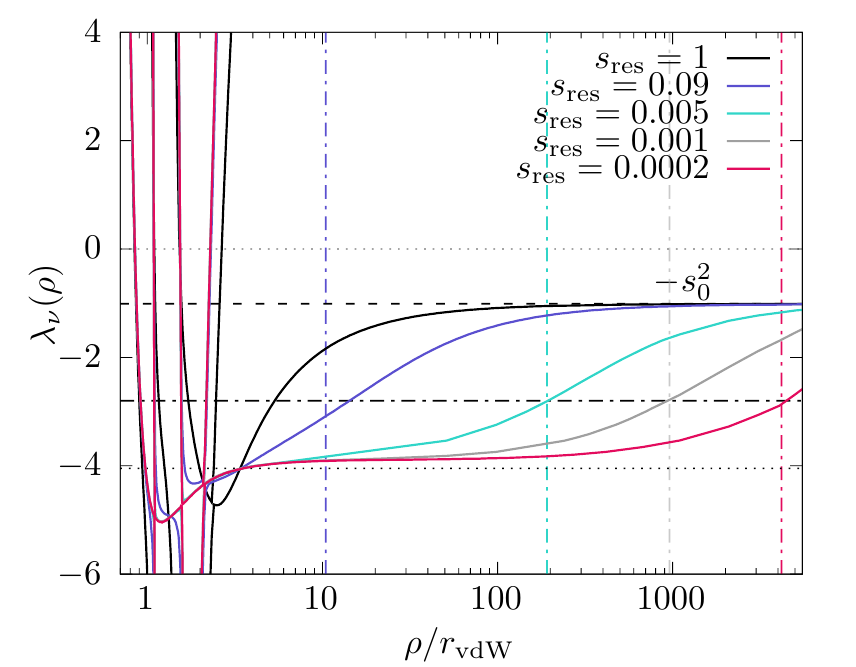}
    \caption{Three-body potentials $\lambda_{\nu}$ ($\nu=-1$ to $1$) for resonances with $s_{\mathrm{res}}$ ranging from 1 to 0.0002. Here, $\lambda_0$ can be seen to converge to $-s_0^2\approx-1.0125$ (the dashed black line) in the scale-free region for each resonance. The vertical dash-dotted lines indicate the value of $r^*$ for the resonance drawn with the same color. The horizontal line indicates $\lambda_0=-2.8$, which is the approximate value for $\lambda_0(r^*)$ for all the narrow resonances.}
    \label{fig:lambda}
\end{figure}

To clearly illustrate that the universal long-range behavior in the resonant regime is governed by $r^*$ in the narrow resonance limit, we have in \cref{fig:lambda} plotted the potentials $\lambda_{\nu}$ for five different resonances with $s_{\mathrm{res}}$ ranging from $1$ to $0.0002$, with $a=\pm\infty$. The additional resonances are obtained from the same potentials by simply varying the coupling strength (see \cref{table:feshbach}). The vertical lines indicate $r^*$ for the intermediate and narrow resonances and the color matches the corresponding potential curves. For the broad and intermediate resonances, i.e., $s_{\mathrm{res}}=1-0.09$, $\lambda_0$ can be seen to converge to the universal constant $-s_0^2$ as the scale-free region for each potential is approached. For the three very narrow resonances ($s_{\mathrm{res}}=0.005-0.0002$) the tails of $\lambda_{0}$ have a similar convergence tendency with respect to $r^*$ as they can be seen to have reached approximately the same height at $r^*$, where $\approx-2.8$ (black dashed-dotted horizontal line) for all three curves. The shape of the tail region is clearly different for $s_{\mathrm{res}}=1$ ($r^*\approx r_{\mathrm{vdW}}$) and the curve for the intermediate resonance looks more like the ones in the narrow resonance regime.

\subsubsection{Avoided crossings and resonance strength}

Prompted by the difference in sharpness of the avoided level crossings, we now examine in more detail how the resonance strength affects the energy gaps between the non-crossing curves $U_{\nu}$.
\begin{figure*}[htbp]
    \centering
    \includegraphics{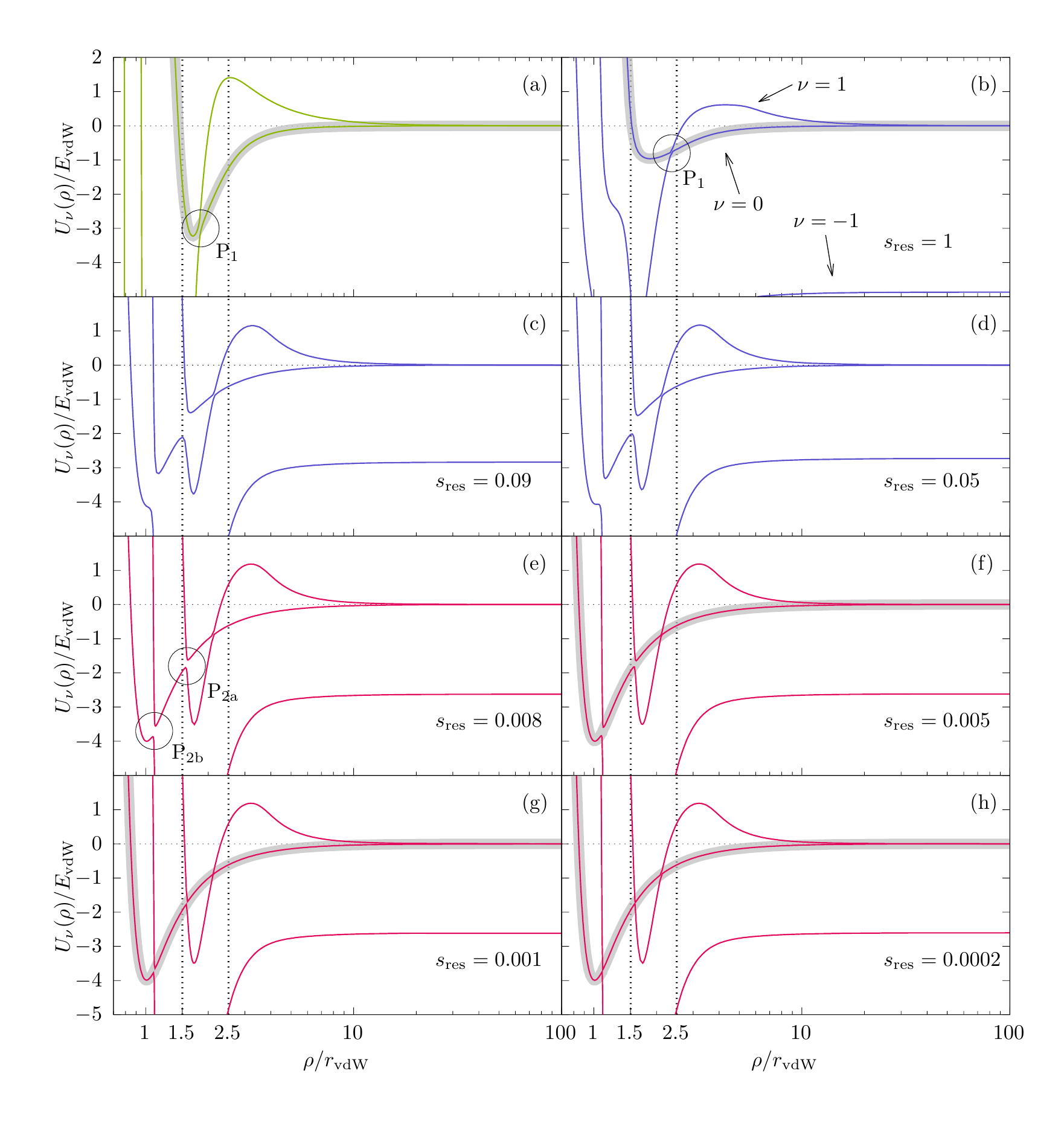}
    \caption{Three-body adiabatic potential curves $U_{\nu}$ for $a \rightarrow \pm\infty$ calculated using a single channel model with one $J=0$ two-body level ($\nu=-2$ to $1$, in green) and using the two-channel model with different resonance strengths $s_{\mathrm{res}}$ ($\nu=-1$ to $1$, in blue and red).}
    \label{fig:avoidedcrossings}
\end{figure*}
In \cref{fig:avoidedcrossings} we show the three lowest adiabatic potential curves obtained with values of $s_{\mathrm{res}}$ ranging from $1$ to $0.0002$ ((b)--(h)). For comparison, we also include single-channel three-body adiabatic potential curves ($\nu=-2$ to $1$) with one bound $s$-wave in (a). Before discussing the resonance background effects on the adiabatic potential curves, we remark on two features that all the potential energy landscapes have in common. The first feature is a peak reminiscent of that of a centrifugal barrier (this is a broad avoided crossing with a higher-lying channel) and the second is a sharp avoided crossing (marked $\mathrm{P}_1$) of the channel associated with the Efimov effect $\nu=0$ and the lowest three-body scattering channel $\nu=1$. The avoided crossing $\mathrm{P}_1$ appears near $\rho=1.9r_{\mathrm{vdW}}$ in the single-channel energy landscape and close to $\rho=2.4r_{\mathrm{vdW}}$ in the two-channel energy landscape for $s_{\mathrm{res}}=1$ (b). The position of $\mathrm{P}_1$ can be seen to slightly depend on the resonance strength as it is shifted to smaller $\rho$ as $s_{\mathrm{res}}$ decreases ($\rho=2.1r_{\mathrm{vdW}}$ for $s_{\mathrm{res}}=0.0002$ (h)). The energy gap of $\mathrm{P}_1$ is, however, static with respect to the resonance strength.

The potential energy landscape changes as the Feshbach resonance is tuned to the narrow resonance regime. As $s_{\mathrm{res}}$ is decreased, the peak of the centrifugal barrier rises in energy, and two new avoided crossings gradually appear in the spectrum (marked $\mathrm{P}_{2\mathrm{a}}$ and $\mathrm{P}_{2\mathrm{b}}$ in (e)). Apparently, there are two kinds of avoided crossings in our spectra, firstly a static kind, $\mathrm{P}_1$,  which is independent of the channel couplings, and secondly a coupling-dependent kind, $\mathrm{P}_{2}$. Our interpretation is that the avoided crossing $\mathrm{P}_1$ originates from the geometric deformation of the three-body state that was discussed in \cref{sec:Prelude} and that the two avoided crossings $\mathrm{P}_{2\mathrm{a}}$  and $\mathrm{P}_{2\mathrm{b}}$ are between states belonging to different hyperfine channels and hence connected only by spin-flip transitions. 

Nonadiabatic transitions (i.e., jumping the gap, or following the diabatic path) take place close to the center\footnote{The point where two diabatic asymptotes to the potential curves cross.} of an avoided crossing. Trajectories that pass through the apex of a cusp\footnote{This is the case where the two avoiding potential curves have sharp cusps separated by a small energy gap. The situation is realized for all three avoided crossings in \cref{fig:avoidedcrossings}(h).} at an avoided crossing transition with almost unit probability, whereas the wave bifurcates if the trajectory that transverses the potential energy curves passes close to its center. One of the waves then continues along the adiabatic path and the other one tunnels through the avoided crossing to the other adiabatic potential energy curve with a probability that can be calculated using the Landau$-$Zener formula \cite{Landau,Zener}. If the avoided crossing region is sufficiently small, the nonadiabatic transition probability can be expressed as 
\begin{equation}\label{eq:LZ}
    P_{\mathrm{LZ}}=\exp{{\frac{-\pi^2}{h}\frac{g^2}{v \delta s}}},
\end{equation}
where $g$ is the energy gap, $v$ is the local velocity and $\delta s$ is the difference of the slopes of the two crossing diabatic asymptotes that trace through the center \cite{Devaquet}. 


Possible diabatic curves asymptotically connecting to the Efimov potential are indicated in gray in \cref{fig:avoidedcrossings}. In the limit of very narrow resonances this is readily determined by comparing to the extreme case (h) where the coupling approaches zero. In the opposite limit the proper diabatization can be inferred by comparing to the familiar single-channel model. That is, for the one-channel case (a) and for broad resonances (b) the diabatic potential bridges the gap P1, which, as will be discussed below gives rise to the familiar van der Waals universality. For very narrow resonances there also seems to be a universal behavior, though qualitatively different in that the most natural diabatization in addition to P1 also bridges the gaps P2. The intermediate resonances do not have an easily recognizable diabatization. The topologically most reasonable path, which also has the highest probability, trace through $\mathrm{P}_1$ together with $\mathrm{P}_{2\mathrm{a}}$ and $\mathrm{P}_{2\mathrm{b}}$.

Moreover, the Efimov potential in (h) is associated with the closed channel, while the two red curves are identical to the background potential, i.e., the hyperradial potentials for $V_{\rm bg}$ (with $a_{\rm bg}=63 a_0$) in the absence of any resonant couplings. The Efimov potential gradually acquires more of the open channel character as $\rho$ becomes very large, which is indicated by the convergence of the tail to the universal Efimov attraction \cref{eq:efimov_attraction}.

\subsubsection{Three-body effective potentials}
\begin{figure}[t]
    \centering
    \includegraphics{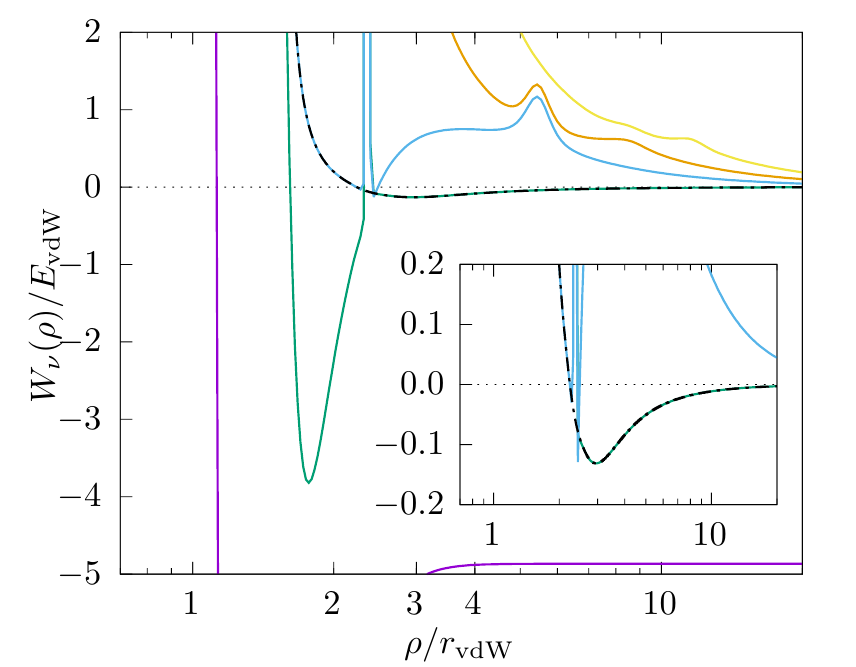}
    \caption{Effective potential curves $W_{\nu}$ ($\nu=-1$ to $3$) for $s_{\mathrm{res}}=1$, together with the diabatized effective potentials (black dashed-dotted curves). The Efimovian diabatic potential here traces through a single sharp peak. Note that the energy scaling on the vertical axis is the same as that in \cref{fig:avoidedcrossings} and the universal Efimov attraction is the black dashed curve.}
    \label{fig:Wbroad}
\end{figure}

\begin{figure*}[t]
    \centering
    \includegraphics{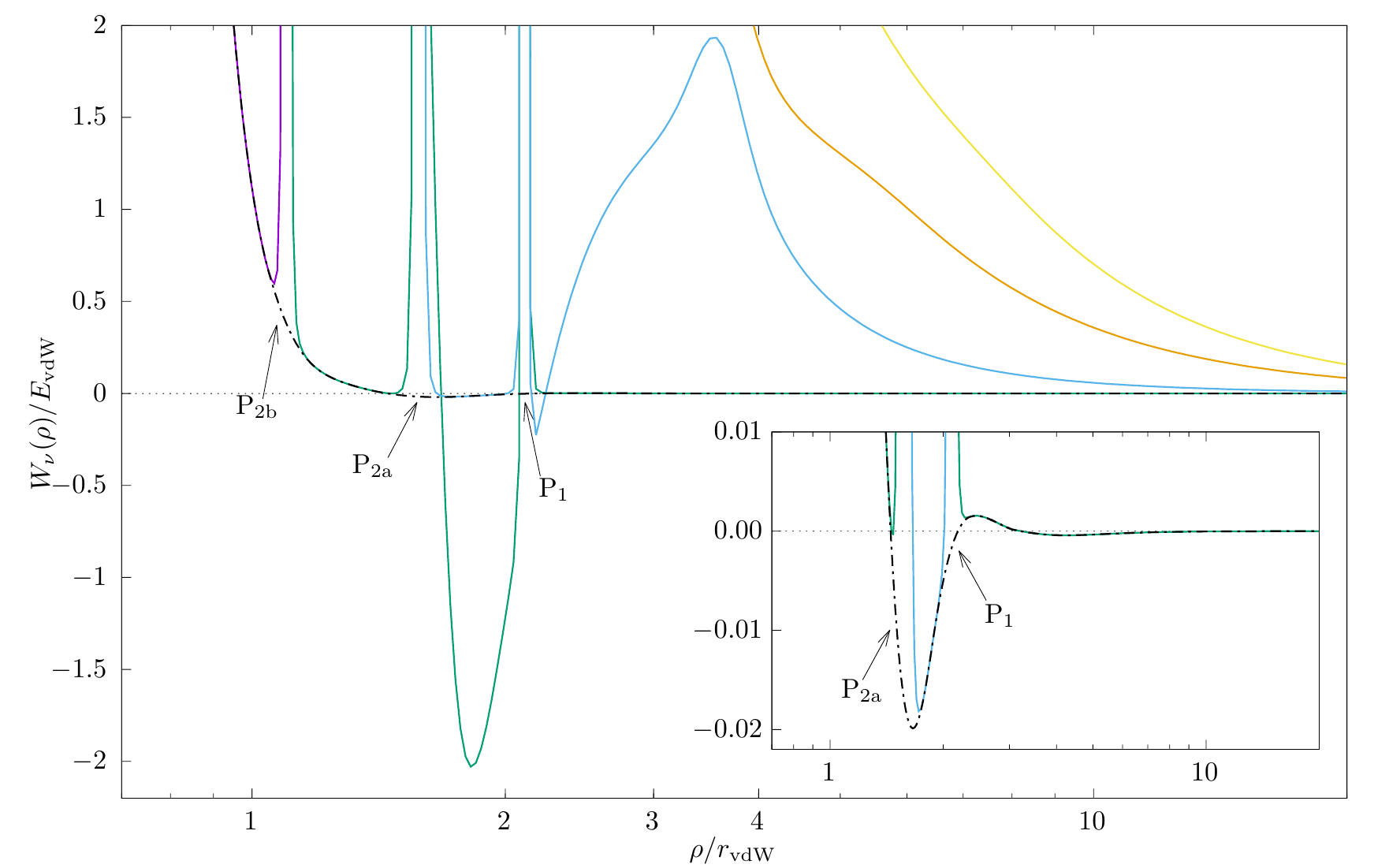}
    \caption{Effective potential curves $W_{\nu}$ ($\nu=-1$ to $3$) for $s_{\mathrm{res}}=0.0002$. The Efimovian diabatic potential here traces through three sharp peaks.}
    \label{fig:Wnarrow}
\end{figure*}

To illustrate the effect of adding the diagonal correction to the adiabatic potentials, we show in \cref{fig:Wbroad,fig:Wnarrow} the resulting three-body effective potentials $W_\nu$ ($\nu=-1$ to $3$) in the limit of the broad resonance regime and for the narrow resonance, i.e., with $s_{\mathrm{res}}=1$ and $s_{\mathrm{res}}=0.0002$. For 
 $s_{\mathrm{res}}=1$, in \cref{fig:Wbroad}, a single sharp peak can be observed at the location of the avoided crossing $\mathrm{P}_1$ in the corresponding adiabatic curves. For $s_{\mathrm{res}}=0.0002$, in \cref{fig:Wnarrow}, three sharp peaks with maxima at the location of the three avoided crossings $\mathrm{P}_1$, $\mathrm{P}_{2\mathrm{a}}$ and $\mathrm{P}_{2\mathrm{b}}$ are visible. The sharp peaks are artefacts of the adiabatic representation. While the diabatic states do not drastically change their character at the avoided crossings, the adiabatic states are associated with different diabatic potentials on either side of the avoided crossing. Since the crossing is narrow, the adiabatic states will over a short $\rho$-interval change character from one diabatic state to another. This is reflected by a sudden increase of $|Q_{\nu\nu}|$ (\ref{eq:effective_potential}). The diabatic Efimov-potentials suggested in \cref{fig:avoidedcrossings} are indicated as dashed-dotted black curves. In comparison with the adiabatic potentials, the effective diabatic potentials are always shifted upward in energy as compared to the adiabatic potentials. The effect of adding the nonadiabatic correction is most dramatic for the potential associated with the Efimov effect, where the attractive well becomes very shallow compared to the single-channel model.

\begin{figure}[ht]
    \centering
    \includegraphics{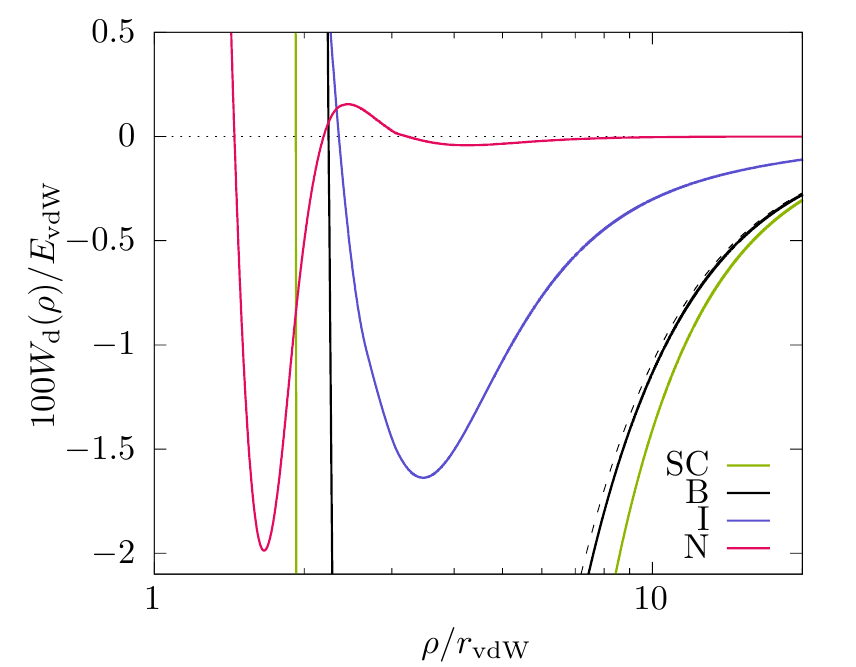}
    \caption{Diabatic Efimovian effective potential curves $W_{\mathrm{d}}$ in the broad ($s_{\mathrm{res}}=1$), intermediate ($s_{\mathrm{res}}=0.09$) and narrow ($s_{\mathrm{res}}=0.0002$) regimes, together with the effective potential from the single-channel model with one $s$-wave bound state. The black dashed curve is the universal Efimov attraction. Note that for the narrow resonance, it is the very shallow well outside the local maximum which corresponds to the Efimov-like attraction.}
    \label{fig:Wzoom}
\end{figure}

In \cref{fig:Wzoom} we zoom in close to the zero energy threshold and show the diabatic Efimovian effective potential curves for three resonances in the broad, intermediate, and narrow regime together with the single-channel potential with one $s$-wave bound state. For the hyperradial range covered in the figure, only the potential for the broadest resonance ($s_{\mathrm{res}}=1$, $r_{\mathrm{c}}=r_{\mathrm{vdW}}$) has a tail that is within its scale-free region. This potential can indeed be seen to converge to the universal Efimov attraction (black dashed curve) close to $\rho\approx 10r_{\mathrm{vdw}}$. Note also that for $s_{\mathrm{res}}=1$ the convergence to the universal attraction occurs at even smaller $\rho$ than for the single-channel potential. The potentials for the broad and intermediate resonances show $s_{\mathrm{res}}$-dependent behavior similar to that observed in \cite{Kraats_Kookelmans}. The position of the repulsive wall, which is located at $\rho\approx2.2r_{vdW}$ for $s_{\mathrm{res}}=1$, shifts to slightly larger $\rho$ and the depth of the potential decreases as $s_{\mathrm{res}}$ is decreased. The potential for $s_{\mathrm{res}}=0.0002$ stands out from the other two since it has two minima and two repulsive regions. Tracing inwards from large $\rho$, a shallow well appears, the potential then crosses zero at $\rho\approx3.2r_{\mathrm{vdw}}$ and has a centrifugal barrier-like peak, the potential again becomes attractive and a second deeper well appears. The potential finally becomes repulsive again at $\rho\approx1.5r_{\mathrm{vdw}}$. The inner potential well is the topological result of a combination of the connections through the avoided crossings $\mathrm{P}_1$ and $\mathrm{P}_{2\mathrm{a}}$. Since the first well is the one that could possibly contain Efimov states, it is reasonable to compare the first repulsive part of this potential with the ones previously discussed. The trend then dictates that the hyperradius of the zero crossing position, i.e., the position of the repulsive wall, slightly increases, and the depth of the potential minima decreases with decreasing $s_{\mathrm{res}}$.

\subsection{Summary}

To summarize our results: we have examined beyond single-channel effects on the three-body potentials and specifically focused on how the resonance strength affects the position of the repulsive wall and the depth of the attractive well in the Efimovian effective potential. We found that the aforementioned properties of the effective potentials are related to two kinds of avoided crossings in the adiabatic energy landscapes. The first kind, labeled $P_1$ in \cref{fig:avoidedcrossings}, occurs close to the three-body deformation hyperradius where the Efimovian orbital is forced to transform \cite{NaidonPascalUeda}. For broad to intermediate resonances, as in single-channel models, bridging the gap of this avoided crossing gives rise to the repulsive wall in the Efimovian effective potential. Tuning the Feshbach resonance strength from the broad to the intermediate regime moves the repulsive wall outwards, consistent with the observation in \cite{Kraats_Kookelmans}. In the narrow limit, the potential acquires a second inner well and hence the wall turns into a barrier. For the narrow resonance, the repulsive barrier is formed outside the avoided crossing region and a diabatization through $P_1$ and $P_{2\mathrm{a}}$ forms the inner well and the second zero-energy crossing of the effective potential.

\section{Conclusions}

We have developed a method to explicitly include the multichannel-nature of physics at Feshbach resonances into the Schr\"{o}dinger equation for three-body systems expressed in the hyperspherical adiabatic equation. Using this model we can study effects arising from the finite width of the resonance. Our calculations have focused on narrow to intermediate-strength resonances in the $^{23}{\mathrm{Na}}$, but can easily be applied to any other system.  

In this paper, we analyzed the Efimovian effective potential and found that the position of the repulsive wall, the depth of the energy minimum, and tail convergence are clearly dependent on the resonance strength. As $s_{\mathrm{res}}$ is decreased we observe an increase in the hyperradius where the potential associated with the Efimov effect crosses zero energy, and correspondingly a decrease in well depth. Consequently, the onset of the characteristic Efimov trimer energy spectrum is pushed to very weak binding energies, and correspondingly large length scales, which makes them difficult to detect in experiments.   

The van der Waals universality is an effect of the universal form of the attractive well formed in the Efimovian hypespherical potential. This in turn is a result of the interplay between the position of the repulsive wall, which always appears close to the deformation hyperradius, and the long-range convergence to the universal Efimov attraction, which \emph{is} dependent on the intrinsic length of the Feshbach resonance. Thus it is evident that potentials governed by $r^*$ deviate from the universal intermediate-region hyperradial form characteristic for potentials displaying van der Waals universality. Therefore, it is necessarily true that the van der Waals universality breaks down in the narrow resonance limit.

All results in this paper are for an infinite scattering length, and limited to the study of the potentials. We expect that in the case of finite $a$, it is only possible for the Efimovian potential to contain trimer states with the proper universal scaling if $|a|\gg r^*$ and $|a|\gg r{\rm vdW}$. The size of the first trimer state with universal scaling must then be in parity with $\rho$ at the onset of the scale-free region and thus proportional to the resonance intrinsic length. Since the Efimov potential has an energy well with a minimum that appears in the region $\rho\sim 3-4 r_{\mathrm{vdW}}$ for all the examined resonances, there could also exist lower energy trimers with a ground state size determined by the equilibrium hyperradius and possibly also higher levels which do not exhibit the universal geometric scaling property. This will be the subject of future studies. 

 \section*{Acknowledgements}

We gratefully acknowledge support from the Swedish Research Council (VR), grant 2017-03822.

The computations were enabled by resources provided by the National Academic Infrastructure for Supercomputing in Sweden (NAISS) and the Swedish National Infrastructure for Computing (SNIC) at the PDC Center for High Performance Computing, KTH Royal Institute of Technology and at the High Performance Computing Center North (HPC2N), Umeå University, partially funded by the Swedish Research Council through grant agreements no. 2022-06725 and no. 2018-05973.

\clearpage


\begin{widetext}

\begin{table}[th]
	\centering
	\caption{The two top rows show the physical properties of two Feshbach resonances found in ultracold scattering experiments with $^{23}\mathrm{Na}$ atoms \cite{Chin_2010} and the parameters used to model these resonances in the three-body calculations. The difference in magnetic moment $\delta\mu/\mu_{\mathrm{B}}$, where $\mu_{\mathrm{B}}$ is the Bohr magneton. The bottom rows show the input used for generating Feshbach resonances with varying $s_{\mathrm{res}}$.}
	\label{table:feshbach}
	\begin{tabular*}{1.\linewidth}{@{\extracolsep{\fill}}lllllll}
		\hline
		\hline
		\multicolumn{4}{l}{Physical properties}& \multicolumn{3}{l}{Model parameters}\TBstrut\\
		\hline
		$B_0$(mT)&$\Delta B$(mT)&
  $s_{\mathrm{res}}$&$r^*/r_{\mathrm{vdW}}$
  &$B_{\mathrm{c}}$(mT)&
  $W_0\times10^{6}$(a.u.)&$B_0^*$(mT)\TBstrut\\
		\hline
		$90.7$&$0.1$&0.09&10.4&$90.75620$&$49.6228$&$90.71084$\Tstrut\\
		$85.3$&$2.5\times10^{-4}$&0.0002&4170&$85.30014$&$2.48114$&$85.30003$\Bstrut\\
		\hline
		$85.3$&$1.0905\times10^{-3}$&0.001&956&$85.30061$&$5.18196$&$85.30012$\Tstrut\\
		$85.3$&$5.4525\times10^{-3}$&0.005&191&$85.30306$&$11.5872$&$85.30059$\\
		$90.7$&$8.7240\times10^{-3}$&0.008&119&$90.70490$&$14.6568$&$90.70089$\\
		$90.7$&$1.0905$&1&0.956&$91.31291$&$163.868$&$90.83035$\Bstrut\\
		\hline\hline
	\end{tabular*}
\end{table}

\end{widetext}

\bibliography{bibliography.bib}

\end{document}